\documentclass[aip,reprint]{revtex4-1}
\usepackage{graphicx}
\usepackage{dcolumn}
\usepackage{bm}

\usepackage[utf8]{inputenc}
\usepackage[T1]{fontenc}
\usepackage{mathptmx}
\usepackage{etoolbox}
\usepackage{xcolor}

\draft 

\begin{document}

\title{Fourier-transform spectroscopy and global deperturbation treatment of the $A^1\Sigma^+_u$ and $b^3\Pi_u$ states of K$_2$ in the entire bound energy range}

\author{I. Klincare}
\author{A. Lapins}
\author{M. Tamanis}
\author{R. Ferber}
\email{ferber@latnet.lv}
\affiliation{Laser Center, Faculty of Physics, Mathematics and Optometry, University of Latvia, 19 Rainis blvd, Riga LV-1586, Latvia}

\author{A. Zaitsevskii}
\affiliation{Department of Chemistry, Lomonosov Moscow State University, 119991, Moscow, Leninskie gory 1/3, Russia}
\author{E.A. Pazyuk}
\author{A.V. Stolyarov}
\email{avstol@phys.chem.msu.ru}
\affiliation{Department of Chemistry, Lomonosov Moscow State University, 119991, Moscow, Leninskie gory 1/3, Russia}

\date{\today}

\begin{abstract} 

Rotationally-resolved Fourier-transform spectra of laser-induced fluorescence (LIF) $A^1\Sigma^+_u\sim b^3\Pi_u \rightarrow X^1\Sigma^+_g$ of K$_2$ molecules were recorded and analyzed yielding 4053 term values of spin-orbit (SO) coupled $A \sim b$ complex of $^{39}$K$_2$ isotopologue  with ca 0.01 cm$^{-1}$ accuracy. Their compilation with 1739 term values from previously published sources allowed to cover the energy range [9955, 17436] cm$^{-1}$ from the bottom of the lower-lying $b^3\Pi_u$ state up to the vicinity of the atomic asymptote $4s\;^2{\rm S}_{\frac{1}{2}}$ + $4p\;^2{\rm P}_{\frac{1}{2}}$, with rotational quantum number $J\in[0,149]$. The experimental data were processed by a direct 6$\times$6 coupled-channel (CC) deperturbation treatment, which accounted explicitly for both SO and electronic-rotational interactions between all six $e$-symmetry states: $A^1\Sigma^+_u$($0^+_u$), $b^3\Pi_u$($0^+_u$,$1_u$,$2_u$), $c^3\Sigma_u$($1_u$), and $B^1\Pi_u$($1_u$). The initial parameters of the global deperturbation model have been estimated in the framework of {\it ab initio} electronic structure calculations applying multi-referenced configuration-interaction and coupled-clusters methods. The interatomic potentials analytically defined for $A$ and $b$ states, as well as SO-splitting of the triplet $b$ state and $A\sim b$ SO-coupling functions have been particularly refined to fit 5792 term values of the $^{39}$K$_2$ isotopologue, whereas the rest parameters were fixed on their {\it ab initio} values. The resulting mass-invariant parameters of the 6$\times$6 CC model reproduced the overall rovibronic term energies of the $A \sim b$ complex of $^{39}$K$_2$ with the accuracy, which is well within the experimental errors. The quality of the deperturbation analysis was independently confirmed by comparison with the present obtained 705 and 14 term values of respective $^{39}$K$^{41}$K and $^{41}$K$_2$ isotopologues, as well as by agreement between measured and predicted relative intensity distributions in long $A\sim b \rightarrow X(v_X)$ band progressions. This deperturbation analysis provided the refined dissociation energy $T_{dis}$=17474.569(5) cm$^{-1}$ and the long-range coefficient $C_3^{\Sigma}$ = 5.501(4) $\times10^5$cm$^{-1}$\AA$^3$ relevant to the non-relativistic atomic limit $4s$+$4p$. The derived $T_{dis}$ yielded the accurate well depth $D_e$ = 4450.910(5) cm$^{-1}$ for the ground $X^1\Sigma^+_g$ state, whereas the new $C_3^{\Sigma}$ value yielded the improved estimates for atomic K($4p\;^2{\rm P}_{\frac{1}{2};\frac{3}{2}}$) radiative lifetimes, $\tau_{\frac{1}{2}}$=26.67(3) and $\tau_{\frac{3}{2}}$=26.32(3) ns.

\end{abstract}


\maketitle

\section{Introduction}\label{intro}

The low-lying excited $A^1\Sigma^+_u$ and $b^3\Pi_u$ states ($A$ and $b$ for short) of a homonuclear diatomic alkali molecule are attracting a lot of attention for several reasons. First, they open a window to the state manifold of \textit{g}-parity, which is not directly accessible from the ground state of the same parity\cite{Bernath2005book}. In addition, the triplet states become accessible due to an admixture of the $b^3\Pi_u$ state\cite{Field2004book}. More recent interest is connected with producing the stable ultracold molecular ensembles\cite{Julienne2012} since these states can successfully serve as the intermediates to transfer translationally cold molecules produced from ultracold atoms in the weakly bound ground state levels by magneto- or photo-association to the deeply bound ground $X^1\Sigma^+_g$ state~\cite{Tiemann2020}, in particular, to the lowest rovibrational level with $\textit{v}_X$ = $\textit{J}_X$ = 0. 

The challenging issue in studying the singlet $A^1\Sigma^+_u$ and triplet $b^3\Pi_u$ states relates to the spin-orbit (SO) interaction, which rapidly increases as the mass of a dimer increases\cite{Pazyuk15_RCR}. In respect to that, among alkali dimers one can obviously distinguish light (Li$_2$, where the fine structure Li($2p\;^2{\rm P}$) splitting~\cite{NIST} $\Delta E_{fs}=E_{2p^2{\rm P}_{\frac{3}{2}}}-E_{2p^2{\rm P}_{\frac{1}{2}}}$ is small, being 0.336 cm$^{-1}$), intermediate (Na$_2$ and K$_2$, the respective $\Delta E_{fs}$ are 17.1963 cm$^{-1}$ and 57.7103 cm$^{-1}$) and heavy (Rb$_2$ and Cs$_2$, where the respective $\Delta E_{fs}$ values reach 237.595 cm$^{-1}$ and even 554.0393 cm$^{-1}$). 
For Li$_2$ it is often not necessary at all to consider the mutual SO interaction between $A$ and $b$ states in an explicit form~\cite{Linton1996, Russier1997} . The situation becomes completely different for the heaviest Rb$_2$ and Cs$_2$ dimers, in which the $A$ and $b$ states should be considered as a fully SO-mixed $A\sim b$ complex. For these molecules, the 2$\times$2 $A^1\Sigma^+_u(0^+_u) \sim b^3\Pi_u(0^+_u)$ or even 4$\times$4 $A^1\Sigma^+_u(0^+_u)\sim b^3\Pi_u (0^+_u, 1_u, 2_u)$ coupled-channel (CC) deperturbation treatment has unavoidably to be applied yielding the interatomic potential energy curves (PECs) and SO coupling matrix elements allowing to reproduce the measured rovibronic term values of the $A\sim b$ complex with the experimental accuracy, close to 0.01 cm$^{-1}$, see \cite{PRA2013} for Rb$_2$ and \cite{PRA2011, PRA2019} for Cs$_2$. This allows, in particular, to reliably predict transition probabilities and laser frequencies necessary to excite a selected rovibronic level.

The intermediate case of sodium and potassium dimers still leaves some hope for a possibility to separately focus on the singlet $A$ state alone; this simplified approach was used, for instance, by Lyyra et al. \cite{Lyyra1990} and Falke et al. \cite{Lisdat2006} who presented an effective interatomic PEC for the 
SO-deperturbed (more precisely, the SO-decoupled from the $b$ state) $A$ state of K$_2$. The first deperturbative analysis of the $A$ and $b$ states in K$_2$ was accomplished in the framework of the conventional effective Hamiltonian approach (EHA)~\cite{Brown79} by Ross et al. \cite{Ross87}, then by Jong et al. \cite{JongJMS1992} and Lisdat et al. \cite{Lisdat2001}. The similar band-by-band deperturbation treatment of the locally perturbed $A^1\Sigma^+_u$ and $b^3\Pi_u$ states in Na$_2$ was performed using the magnetic Landé factors measured by Hanle effect~\cite {Stolyarov92_Na2, Stolyarov93_Na2}. For Na$_2$ the most accurate 4$\times$4 CC deperturbation analysis of $A$ and $b$ states was performed in a more recent study~\cite{Qi2007}, which accumulates the experimental term values data from 20 sources.

The so far most comprehensive deperturbative study of the coupled $A \sim b$ states in K$_2$ was performed by Manaa et al. \cite{Manaa2002}. The authors had incorporated the experimental term values available at that time, see, \cite{Ross87, Lyyra1990, JongJMS1992, Lisdat2001, AmiotJMS1991, Amiot1995, Kim95}, as well as a considerable amount of the new data~\cite{Manaa2002}. The deperturbative treatment~\cite{Manaa2002} involved 1802 term values and combination differences of $^{39}$K$_2$ isotopologue covering the $A\sim b$ complex region up to about 16~500 cm$^{-1}$, which is, however, still too far from the second relativistic 4$^2$S$_{\frac{1}{2}}$+4$^2$P$_{\frac{1}{2}}$ dissociation threshold being about 17~436 cm$^{-1}$. These data were fitted based on the reduced 4$\times$4 CC model, which takes into account explicitly the SO coupling between the singlet $A$ and triplet $b$ states, as well as the equidistant SO splitting of the triplet state. The unweighted fit yielded an r.m.s. residual of 0.018 cm$^{-1}$ comparable with the accuracy of spectroscopic measurements limited by the Doppler effect. The authors finally presented: (\textit{i}) the fitted set of empirical Dunham coefficients used for Rydberg-Klein-Rees (RKR) potentials construction of $A^1\Sigma^+_u$ and $b^3\Pi_u$ states up to $v_A=88$ and $v_b=93$ vibrational level, respectively, as well as (\textit{ii}) the properly scaled (morphed) diagonal and off-diagonal SO matrix elements obtained due to all-electron multi-reference configuration interaction (MR-CI) electronic structure calculations. The most reliable sets of both non-relativistic ~\cite{Magnier2004} ($^{1,3}\Lambda_{u,g}$) and relativistic~\cite{Magnier2009} ($\Omega_{u,g}^{+/-}$)  {\it ab initio} PECs for singlet and triplet states manifold of K$_2$ have been obtained in the framework of full CI calculations performed for two valence electrons in combination with $l$-dependant core-polarization potentials (CPP-$l$) construction.

The following considerations motivate the present study of the $A$ and $b$ states in K$_2$. As already mentioned\cite{Manaa2002}, over significant energy region the experimental term values are still rather sparse. Also, the data analyzed by Manaa et al. \cite{Manaa2002} involve only the most abundant $^{39}$K$_2$ isotopologue in the limited energy range of the singlet $A$ state ($E_A$<16037 cm$^{-1}$) far from dissociation threshold. We therefore to include into the global deperturbation analysis the extremely accurate term values of the $A\sim b$ complex available up to the 4$^2$S$_{\frac{1}{2}}$+4$^2$P$_{\frac{1}{2}}$ dissociation threshold~\cite{Lisdat2006,Falke2007}. These data mainly belonging to the highly excited rovibronic levels of the $A$ state were obtained by Falke et al.\cite{Falke2007PhD} by applying the sub-Doppler laser absorption spectroscopy of K$_2$ produced in a molecular beam. So, there is hope that addressing these issues may allow providing the global uniform description of the $A^1\Sigma^+_u$ and $b^3\Pi_u$ states of K$_2$, which will reproduce the overall term values set with the experimental (both Doppler and sub-Doppler) spectroscopic accuracy up to the second non-relativistic dissociation limit $4s\;^2{\rm S}+4p\;^2{\rm P}$. To reach this goal, we aimed to considerably enlarge the amount and abundance of experimental term values for $^{39}$K$_2$, as well as to obtain the $A\sim b$ data for the $^{39}$K$^{41}$K and $^{41}$K$_2$ isotopologues. 

\begin{figure}
\includegraphics[scale=0.3]{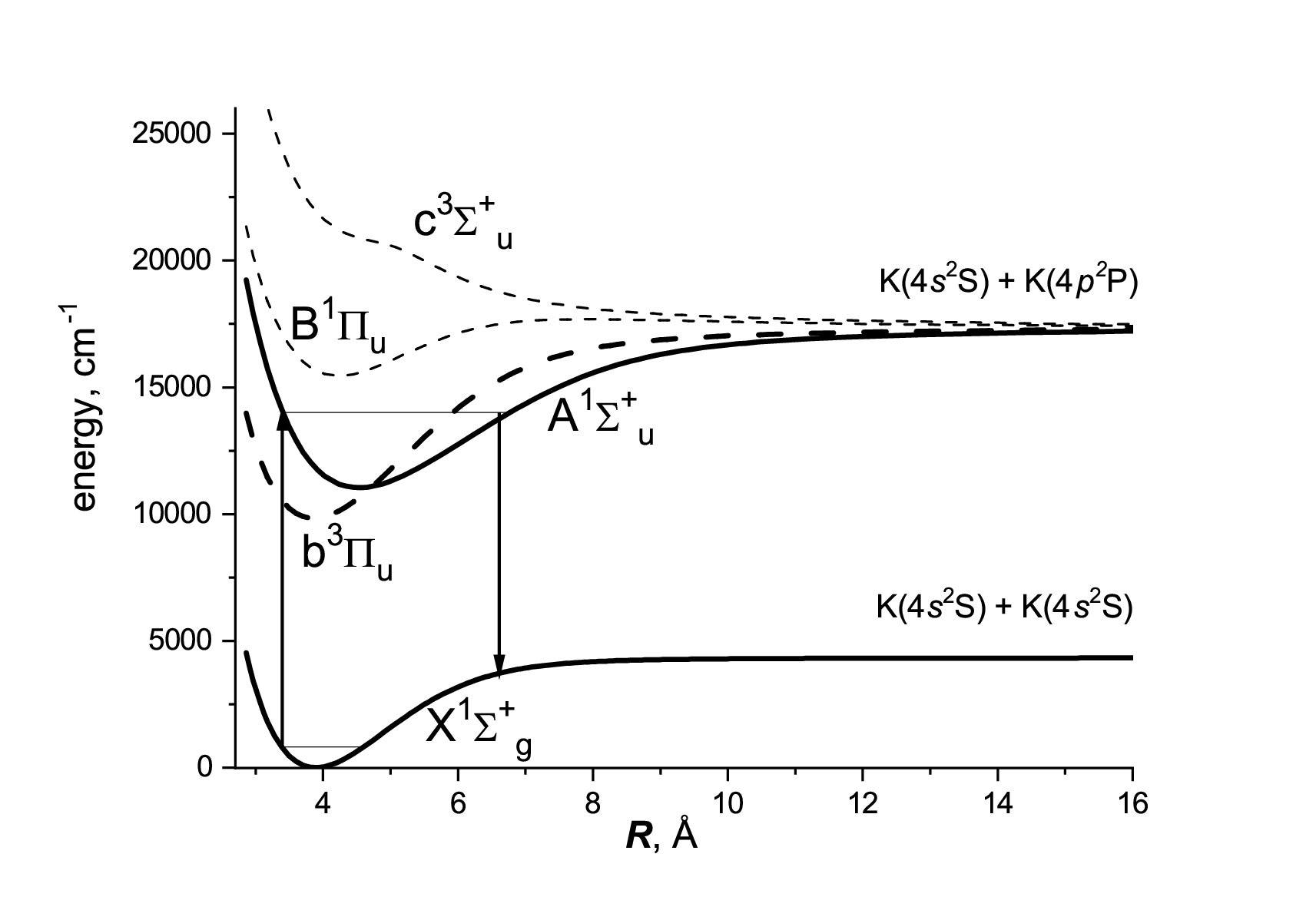}
\caption{The schema of {\it ab initio} scalar-relativistic potential energy curves of the K$_2$ molecule involved into the present consideration.}\label{pec}
\end{figure}

We performed in the University of Latvia Laser Center in Riga the rotationally resolved Fourier-transform (FT) measurements of $A^1\Sigma^+_u\sim  b^3\Pi_u \rightarrow X^1\Sigma^+_g$ fluorescence. The significantly enlarged data field was compiled with the previously published ones, including low-lying\cite{Manaa2002, Kim95} and highly-excited\cite{Lisdat2006, Falke2007} term values. The overall data set has been then treated in the framework of the comprehensive  6$\times$6 CC deperturbation model, which consists of the six $e$-symmetry $\Omega_u=(2)0^+_u,(3)1_u,(1)2_u$ states and takes into account explicitly both SO and electronic-rotational non-adiabatic interactions between four $A^1\Sigma^+_u$, $b^3\Pi_u$, $c^3\Sigma^+_u$, and $B^1\Pi_u$ states, which correspond to pure Hund's "\textbf{a}" case coupling  and converge to the common non-relativistic $4s^2$S+$4p^2$P limit (see Fig.\ref{pec}). The initial structure parameters required as the functions of the internuclear distance $R$, have been estimated in the framework of alternative scalar-relativistic MR-CI method with subsequent account for the SO interactions of scalar states and fully relativistic coupled cluster approach. To clime the mass-invariant property of the derived structure parameters the rovibronic term values predicted for the $^{39}$K$^{41}$K and $^{41}$K$_2$ isotopologues were directly compared with their experimental values. A high reliability of the calculated multi-channel vibrational wavefunctions of the $A\sim b\sim c\sim B$ complex and {\it ab initio} $A-X$ transition dipole moment (TDM) borrowed from Ref.~\cite{Krumins2020} was independently confirmed by a good agreement of the estimated Einstein emission coefficients with the relative intensity distributions measured for the long $A\sim b \rightarrow X(v_X)$ band progressions of $^{39}$K$_2$ and $^{39}$K$^{41}$K isotopologues.


\section{Experiment}

\subsection{Experimental setup}

The rovibrational term values of the $A^1\Sigma^+_u$ and $b^3\Pi_u$ states in K$_2$ were obtained from the FT spectra of the $A\sim b\rightarrow X^1\Sigma^+_g$ laser induced fluorescence (LIF) recorded with the Bruker IFS-125HR spectrometer. The $^{39}$K$_2$, $^{39}$K$^{41}$K, and $^{41}$K$_2$ molecules were produced in a linear heat-pipe oven according to the natural abundance of the stable K isotopes~\cite{Tiecke2019}: 93.26\% of $^{39}$K and 6.73 \% of $^{41}$K. For detection of near infra-red LIF spectra the InGaAs diode was used. About one third of data was picked up from previously recorded FT-LIF spectra during our studies of the $A\sim b$ complex of KCs and KRb molecules~\cite{Kruzins2010, Alps2016}. As is clear, since molecular vapour in the respective heat-pipe operated at about 300$^o$C contained also potassium dimer, the accidentally excited $A\sim b \rightarrow X$ LIF of K$_2$ was also observed and recorded. In the KCs experiments, the tunable home-made external cavity diode lasers with 930, 950, 980, and 1020 nm laser diodes were used for excitation. In the KRb experiments the $A\sim b$ complex of K$_2$ molecules was excited by a Ti:Sapphire laser (Coherent MBR 110). The laser excitation frequencies in these experiments were set within a range 9800--13~000 cm$^{-1}$, which predominantly excited the K$_2$ levels with energy below 13~500 cm$^{-1}$. 

In order to get more systematic term values data on both $A$ and $b$ states of K$_2$ in the higher energy range we performed a new measurement session in the heat-pipe filled only with potassium metal. For this purpose, 10 g of potassium (from \emph{Alfa Aesar}) were heated to about 350$^o$C. The pressure of the buffer gas Ar in the heat-pipe was increased to about 6 mbar to ensure efficient collisional population transfer between rotational levels in the upper excited state. A single mode frequency stabilized Ti:Sapphire laser (Equinox/SolsTis(MSquared)) was exploited, its power at the entrance of the heat-pipe being about 600 mW. Laser frequency was measured with a wavemeter WS7. Backward LIF was recorded with a typical instrumental resolution of 0.03 cm$^{-1}$. The spectral calibration of FT spectrometer has been checked using the Ar and Ne lines from the low current hollow cathode lamps by comparing to the respective data~\cite{NIST, Whaling}.  In order to reach a higher energy range of the $A\sim b$ complex than in our previous measurements, the excitation laser frequency was set within 13~000--13~800 cm$^{-1}$ range, which was the upper limited by the laser performance. Overall about 200 new $A\sim b \rightarrow X$ FT-LIF spectra of K$_2$ were recorded. 

\subsection{LIF spectra analysis}

The recorded $A\sim b (J^{\prime}) \rightarrow X (v^{\prime\prime}, J^{\prime\prime} = J^{\prime}\pm 1)$ emission spectra were processed applying a homemade program, which searches and assigns the rotational lines belonging to a particular LIF progression basing on coincidences between the observed and calculated rovibrational differences. The program operates with the comprehensive ground state term values databases $E^{Calc}_X(v^{\prime\prime}, J^{\prime\prime})$ prepared in advance for all molecules under consideration. The output file reports the following information about the assigned progressions: the molecule (including particular isotopologue), rotational quantum number $J^{\prime}$ of the upper level and its term value $E^{\prime}$, the calculated absorption (excitation) transition, the positions of the calculated and experimental lines assigned to a particular progression, the differences of their positions, the line intensities. In case of K$_2$ the ground $X$ state  term values for $^{39}$K$_2$, $^{39}$K$^{41}$K, and $^{41}$K$_2$ isotopologues have been generated using the properly mass-scaled Dunham's constants borrowed from Ref.\cite{Amiot1995}. The assignment of particular progression was accepted after checking for possible errors, e.g. bad signal-to-noise ratio (SNR), non-realistic absorption transition, etc. Along with the brand-new spectra collection, we have  reanalyzed the FT-LIF spectra previously obtained in KCs and KRb heat-pipes focusing on the K$_2$ progressions.

\begin{figure}
\includegraphics[scale=0.4]{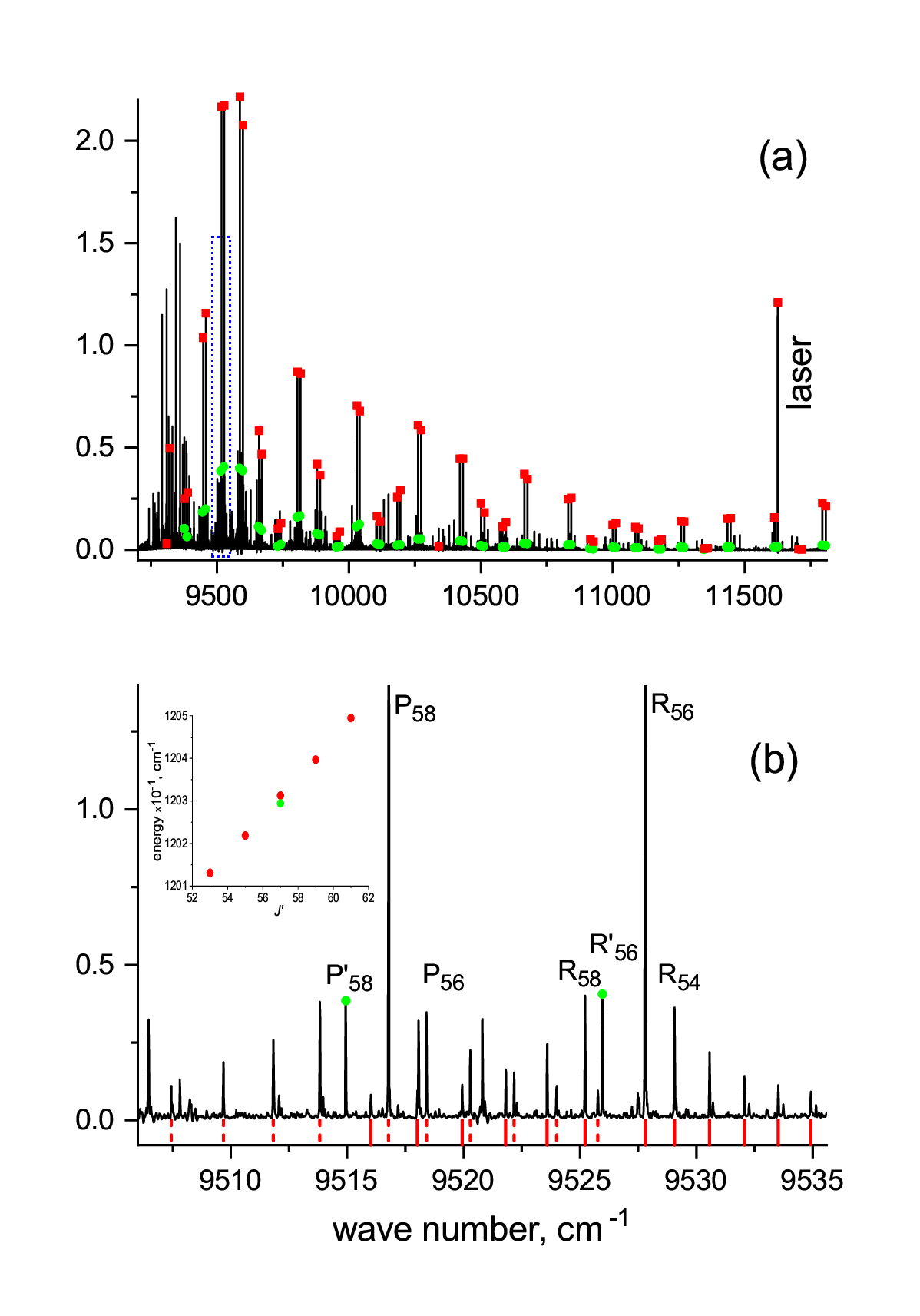}
\caption{The $A^1\Sigma^+_u\sim b^3\Pi_u (J^{\prime})\rightarrow X^1\Sigma^+_g(v_X,J^{\prime\prime}=J^{\prime}\pm 1)$ FT LIF spectrum recorded at excitation laser frequency $\nu_{laser}$ = 11625.540 cm$^{-1}$: (a) full spectrum. Red squares mark $^{39}$K$_2$ doublet $P, R$ progression from the upper state level $E_{A\sim b}(J^{\prime} = 57)$ = 12031.243 cm$^{-1}$ ( $A\sim b$ mixing:  $P_A$ = 67\% and $P_{b0}$ = 31\%). Green points mark the progression from collisionally populated level ($A\sim b$ mixing: $P_A$ = 33\% and $P_{b0}$ = 65\%); (b)  zoomed fragment with assigned satellite lines, see dotted and solid red lines below the spectrum for $P$ and $R$ branches, respectively. The mutual local perturbation of levels with $J^{\prime}$ = 57 is clearly seen in the relaxation lines pattern. The inset presents term values of the respective rovibronic levels as dependent on $J^{\prime}(J^{\prime}+1)$.} \label{spectrum}
\end{figure}

\begin{figure*}
\includegraphics[scale=0.6]{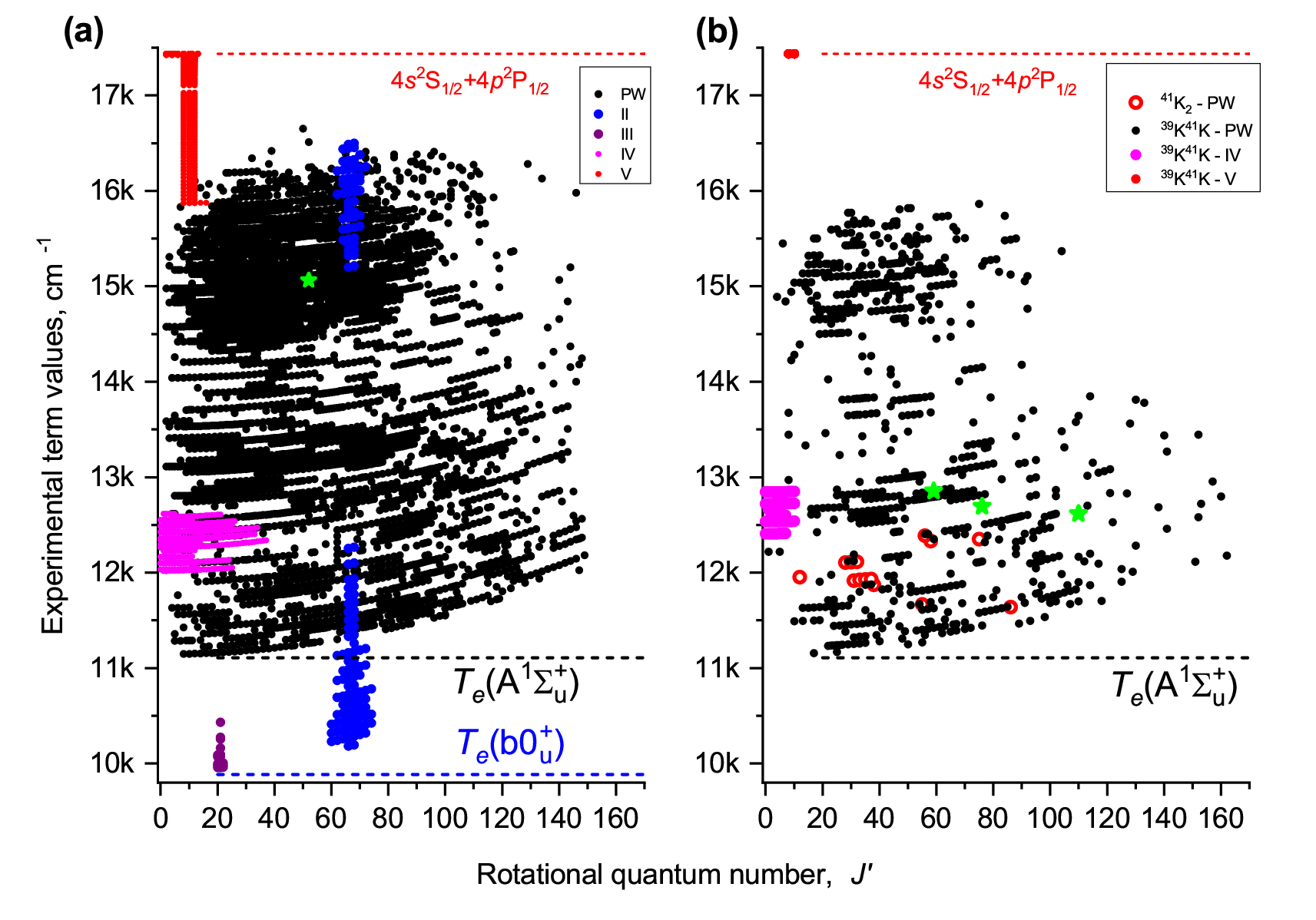}
\caption{Experimental term values of the K$_2$ $A\sim b$ complex as dependent on rotational quantum number $J^{\prime}$: (a) for $^{39}$K$_2$ included in the fit, black points - present work (PW), blue ~\cite{Manaa2002}, magenta ~\cite{Lisdat2001, Lisdat2001PhD}, red ~\cite{Lisdat2006, Falke2007PhD}, purple~\cite{Kim95}; (b) for other isotopologues not included in the fit - $^{39}$K$^{41}$K: black points PW, magenta ~\cite{Lisdat2001, Lisdat2001PhD},  red ~\cite{Falke2007}; for $^{41}$K$_2$: red empty circles PW . Green stars mark the levels, for which relative intensity distributions in LIF progressions are presented, see Sec. \ref{ver}.}
\label{Fig3}
\end{figure*}

In a number of FT-LIF spectra, the satellite lines appeared around strong ("mother's") lines due to depopulation transfer from a directly excited rovibronic level to neighboring rotational levels in collisions with buffer Ar gas atoms. The data processing of satellite lines had provided a great number of additional rovibronic term values of the upper state. Several spectra contained the satellite lines originating from collision-ally populated levels with dominant $b^3\Pi_u(0^+_u)$ character. An example of such spectrum, which was recorded at the KRb study is shown in Fig.~\ref{spectrum}. The full spectrum (Fig.~\ref{spectrum}a) contains several progressions: 3 for K$_2$, 10 for KRb, and 6 for Rb$_2$. The progression marked by blue points originates from the $^{39}$K$_2$ upper level $E_{A \sim b}(J^{\prime} = 57)$ = 12031.243 cm$^{-1}$. A zoomed fragment of spectrum for this progression in the range $v_X$ = 28 is presented in Fig.~\ref{spectrum}b. The additional satellite lines marked as $P^{\prime}$(58) and $R^{\prime}$(56) demonstrate transfer of population to the neighboring $b^3\Pi_u(0^+_u)$ level with $J^{\prime}$ = 57, which lies below the $A^1\Sigma^+_u(0^+_u)$ level with $J^{\prime}$ = 57. Overall more than 6000 term values were obtained. After averaging over energies obtained for the same levels from different spectra the final amount was  4053 term values for $^{39}$K$_2$, see Fig.~\ref{Fig3}a, 705 term values for $^{39}$K$^{41}$K, and 14 term values for $^{41}$K$_2$, see Fig.~\ref{Fig3}b. All present term values are referred to the minimum of the ground state potential by adding the Dunham's correction $Y^X_{00}$~=~-0.022 cm$^{-1}$ from Ref.~\cite{Amiot1995}. 


\section{Direct Deperturbation Fit}\label{DDF}

Direct deperturbation fit (DDF) machinery used for the present treatment of the rotationally resolved $e$-symmetry levels corresponding to the mutually perturbed $A^1\Sigma^+_u$, $b^3\Pi_u$, $c^3\Sigma^+_u$ and $B^1\Pi_u$ states of potassium dimer, which are belonging to the pure Hund's "{\bf{a}}" coupling case ~\cite{Field2004book} and converging to the non-relativistic dissociation threshold (see Fig.~\ref{pec}), consists of the several items described below. 

The developed deperturbation model neglects completely the hyperfine structure (HFS) of the K$_2$ terms since the HFS is not resolved in the Doppler effect limited spectra which were assigned to both $\Omega=0^+_u$-components of the $A\sim b$ complex, singlet $B^1\Pi_u$ state and high rotational levels of both triplet $b^3\Pi_u$ and $c^3\Sigma^+_u$ states. In the Doppler free spectra\cite{Lisdat2001, Lisdat2001PhD, Falke2007, Falke2007PhD} the observed HFS was previously extracted from the measured lines using the HFS version of the effective Hamiltonian approach\cite{Brown2003book}. The deperturbation model can not be formally applied for the mixed $^{39}$K$^{41}$K isotopologue due to a breakdown of the $u \leftrightarrow u$ and $g \leftrightarrow g$ selection rules for intramolecular perturbations. The model tacitly ignores the implicit dependence of the  empirically refined adiabatic interatomic PECs on the reduced molecular mass $\mu$.

\subsection{The molecular Hamiltonian model}\label{Hmodel}

Basically, the vast majority of the observed levels belonging to $A$ and $b$ states can be properly described in the framework of the simplest 2$\times$2 model accounting only for the SO interaction between the $A^1\Sigma^+$ state and the $0^+_u$-component of the triplet $b^3\Pi_u$ state since the higher-lying $B^1\Pi_u$ and $c^3\Sigma^+_u$ electronic terms do not cross the $A$ and $b$ states at any internuclear distance $R$. Nevertheless, the $B$ and $c$ states were still involved explicitly into the present consideration in order to account precisely for the SO-coupling effect on the quasi-degenerate levels located in a vicinity of the common non-relativistic dissociation limit. For intermediate and high rotational levels the diagonal SO-splitting of the triplet $b$ state and spin-rotational interaction between its components $b^3\Pi(0^+_u,1_u,2_u)$ should be taken into account leading to the 4$\times$4 deperturbation model corresponding to the intermediate  between  Hund's "{\bf{a}}" and "{\bf{b}}" coupling cases\cite{Field2004book}. 

The non-adiabatic eigenvalues $E^{CC}_j(J)$ and eigenfunctions $\Phi^{CC}_j(R)$ corresponding to the $j$-level of the $A\sim b\sim c\sim B$ complex of the K$_2$ molecule were described within the framework of the 6$\times$6 CC deperturbation model based on a numerical solution of the six coupled radial Schrödinger equations
\begin{eqnarray}\label{CC}
\left(- {\bf I}\frac{\hbar^2 d^2}{2\mu dR^2} + {\bf V}(R;\mu,J) - {\bf I}E^{CC}_j(J)\right)\Phi^{CC}_j(R) = 0
\end{eqnarray}
with the conventional boundary $\Phi^{CC}_j(0)=\Phi^{CC}_j(+\infty)=0$ and normalization $\sum_iP_i=1$ conditions, where $i\in[A^1\Sigma^+_u;b^3\Pi_{0^+_u,1_u,2_u};c^3\Sigma^+_{1_u};B^1\Pi_u]$ is the index of electronic $\Omega$-state under the consideration, namely: $\Omega\in [(2)0^+_u,(3)1^+_u,(1)2^+_u]$. ${\bf I}$ is the identity matrix and $P_i=\langle\phi_i|\phi_i\rangle$ is the fractional partition of the multi-channel wavefunction $\Phi^{CC}_j$. 

Hereafter, the ${\bf V}(R;\mu,J)$ is the symmetric 6$\times$6 matrix of potential energy having the following diagonal
\begin{eqnarray}\label{Ham_diag}
\langle A^{1}\Sigma^{+}|H|A^{1}\Sigma^{+}\rangle & = & U_A + B(X+2)\\
\langle b^{3}\Pi_{0^+}|H|b^{3}\Pi_{0^+}\rangle & = & U_{b0} + B(X+2)\nonumber \\
\langle b^{3}\Pi_{1}|H|b^{3}\Pi_{1}\rangle & = & U_{b0} + A_{so} + B(X+2)\nonumber \\
\langle b^{3}\Pi_{2}|H|b^{3}\Pi_{2}\rangle & = & U_{b0} + 2A_{so} + B(X-2)\nonumber \\
\langle B^{1}\Pi|H|B^{1}\Pi\rangle & = & U_B + BX\nonumber \\
\langle c^{3}\Sigma^{+}_{1}|H|c^{3}\Sigma^{+}_{1}\rangle & = & U_c + B(X+2)\nonumber
\end{eqnarray}
and the non-vanishing off-diagonal matrix elements
\begin{eqnarray}\label{Ham_offdiag1}
\langle A^{1}\Sigma^{+}|H|b^{3}\Pi_{0^+} \rangle & = & - \sqrt{2}\xi_{Ab0} \\
\langle A^{1}\Sigma^{+}|H|b^{3}\Pi_{1} \rangle & = & - \eta_{Ab1} B\sqrt{2X}\nonumber \\
\langle b^{3}\Pi_{0}|H|b^{3}\Pi_{1} \rangle & = & - B(1-\gamma_{bb})\sqrt{2X}\nonumber \\
\langle b^{3}\Pi_{1}|H|b^{3}\Pi_{2} \rangle & = & - B(1-\gamma_{bb})\sqrt{2(X-2)}\nonumber
\end{eqnarray}
\begin{eqnarray}\label{Ham_offdiag2}
\langle B^{1}\Pi|H|b^{3}\Pi_{1}\rangle & = & -\xi_{Bb}\\
\langle B^{1}\Pi|H|c^{3}\Sigma^{+}_{1}\rangle & = & \xi_{Bc} \nonumber\\
\langle b^{3}\Pi_{1}|H|c^{3}\Sigma^{+}_{1}\rangle & = & \xi_{bc} \nonumber
\end{eqnarray}
\begin{eqnarray}\label{Ham_offdiag3}
\langle A^{1}\Sigma^{+}|H|B^{1}\Pi \rangle & = & - BL_{AB}\sqrt{2X}\\
\langle b^{3}\Pi_{0^+}|H|c^{3}\Sigma^{+}_{1} \rangle & = & - BL_{bc}\sqrt{X}\nonumber\\
\langle b^{3}\Pi_{2}|H|c^{3}\Sigma^{+}_{1} \rangle & = & - BL_{bc}\sqrt{(X-2)},\nonumber
\end{eqnarray}
where 
\begin{eqnarray}\label{BR}
X \equiv J(J+1);\quad B \equiv \frac{\hbar^2}{2\mu R^2}.
\end{eqnarray}

All electronic structure parameters involved in the model are assumed to be the mass-invariant functions of the internuclear distance $R$. In particular, $U_i(R)$ in Eqs.(\ref{Ham_diag}) are the effective (SO-deperturbed) interatomic potentials for the $A^1\Sigma^+_u$, $b^3\Pi_{0^+_u}$,  $c^3\Sigma^+_{1_u}$ and $B^1\Pi_u$ states, $A_{so}(R)$ is the on-diagonal SO splitting function of the triplet $b$ state. The $\xi_{ij}(R)$ in Eq.(\ref{Ham_offdiag2}) and $L_{ij}(R)$ in Eq.(\ref{Ham_offdiag3}) functions are the off-diagonal SO coupling and $L$-uncoupling electronic matrix elements, respectively. The dimensionless parameters $\eta_{Ab1}$ and $\gamma_{bb}$ from Eq.(\ref{Ham_offdiag1}) are considered here as a small constants arisen from the 2-nd order SO$\times$$L$-uncoupling interaction with the remote singlet and triplet $u$-states manifold (excluding a contributions of $B^1\Pi_u$ and $c^3\Sigma^+_u$ states, respectively):  
\begin{eqnarray}\label{2nd_Ham}
\eta_{Ab1} \approx \sum_{j\in ^1\Pi_u} \frac{L_{Aj}\xi_{jb}}{(U_A+U_b)/2-U_j};\quad 
\eta_{bb}\approx \sum_{j\in ^3\Sigma^+_u} \frac{L_{bj}\xi_{jb}}{U_b-U_j}.
\end{eqnarray}

\subsection{{\it Ab initio} electronic structure calculations} \label{abinitio}

Interatomic potentials and non-adiabatic electronic matrix elements involved into the applied deperturbation 6$\times$6 CC model are obviously needed to be estimated in a wide range of $R$ since we intend to deperturb the $A\sim b\sim c\sim B$ complex in the entire boundary range starting from the bottom of the $b$ state and ending on its common non-relativistic $4s\;^2{\rm S}$ + $4p\;^2{\rm P}$ dissociation limit. It has been done below by means of alternative scalar-relativistic and fully-relativistic electronic calculations accomplished by multi-referenced configuration-interaction and coupled-clusters methods, respectively.

\subsubsection{Scalar-relativistic approximation $-$ Hund's "{\bf{a}}" coupling case}\label{CICPP}

The scalar-relativistic (sr) adiabatic PECs $U^{sr}_i(R)$ and respective electronic eigenfunctions, $\Psi^{sr}_i(\textbf{r},R)$, corresponding to a pure Hund's "{\bf{a}}" coupling case were obtained within the framework of the two-valence electrons configuration interaction (CI) calculations combined with the model treatment of core-valence interaction by means of core polarization potentials (CPPs) \cite{Magnier2004, Magnier2009, CPP}. The computational details of the CI-CPP procedure can be found in Refs~\cite{Pazyuk2015, Pazyuk2016} on the heavier Rb$_2$ and Cs$_2$ dimers. Briefly, the atomic core of potassium was replaced by the averaged relativistic ECP10MDF effective core potentials~\cite{Lim2005, Sadlej1991}, leaving 1 valence and 8 sub-valence electrons of each atom for explicit treatment. The molecular orbitals (MO) were generated using the state-averaged complete active space self-consistent field (SA-CASSCF) method~\cite{Werner85}, taking the (1-5)$^{1,3}\Sigma^+_{u/g}$, (1-3)$^{1,3}\Pi_{u/g}$, and (1)$^{1,3}\Delta_{u/g}$ states with equal weights. The dynamic correlation was treated explicitly only for the two valence electrons within the multi-reference configuration interaction (MR-CISD) calculations~ \cite{Knowles92}. At both CASSCF and CI steps all (eight) sub-valence MOs were kept doubly occupied. The rest core-polarization and core-valence correlations were taken into account using the $l$-independent CPPs~\cite{CPP} which the cutoff parameter was adjusted to represent exactly the experimental energy $\nu^{exp}_{4S-4P}$ of the $4s\;^2{\rm S}\to 4p\;^2{\rm P}$ transition in K atom. All scalar-relativistic calculations were conducted using the MOLPRO package~\cite{MOLPRO}. 

\subsubsection{Fully relativistic model $-$ Hund's "{\bf{c}}" coupling case}\label{FSRCC}

Alternatively, adiabatic (avoided crossing) and 
SO-deperturbed (quasidiabatic) interatomic potentials for both $A^1\Sigma^+_u(A0^+_u)$ and $b^3\Pi_u(b0^+_u)$ states of K$_2$ were obtained (together with the off-diagonal $\xi^{rel}_{Ab0}(R)$ SO coupling function and SO-splitting between $b0^+_u$, $b1^+_u$ and $b2^+_u$ states) within the framework of fully relativistic electronic structure calculations corresponding to a pure Hund's "{\bf{c}}" coupling case. For K atom the relativistic (two-component) semilocal shape-consistent effective core potentials (RECP) properly accounting for spin-dependent relativistic effects were used ~\cite{Mosyagin:10a,GRECP}. Eight sub-valence (outer core) electrons, $3s^2\,3p^6$, and one valence electron of each K atom were treated explicitly.
To solve the relativistic 18-electron problem the Fock-space relativistic coupled-cluster (FS-RCC) method \cite{Visscher:01,Eliav:22} has been employed. The Gaussian basis $[7s\,7p\,6d\,4f\,2g]$ adapted to the use with the mentioned ECP was constructed via modification and contraction of the diffuse  primitive functions from Ref.~\cite{Widmark:04}. The Fermi vacuum was defined by the ground-state SCF determinant for the doubly charged molecular ion, so that the neutral states corresponded to the two-particle Fock space sector ($0h2p$). The cluster operator expansion comprised single and double excitations (FS-RCCSD approximation). The model space was spanned by all possible distributions of two valence electrons among 52 Kramers pairs of lowest-energy K$_2^{++}$ spinors. To suppress instabilities caused by intruder states, we employed the dynamic denominator shift technique combined with the extrapolation to the zero-shift limit using matrix Pad\'e approximants~\cite{Zaitsevskii:18a}. The calculations were performed with the appropriately modified DIRAC17 program package~\cite{DIRAC:17,DIRACpaper}.

To transform the resulting adiabatic potentials for both $A0^+_u$ and $b0^+_u$ relativistic states to their 
quasidiabatic SO-deperturbed counterparts corresponding to ``{\bf{a}}'' Hund's coupling case and extract the relevant SO coupling function, we employed the technique based on projecting the solutions of scalar-relativistic (SO-free) problem on the subspace of strongly SO-mixed eigenstates of the full relativistic Hamiltonian ~\cite{Zaitsevskii:17}. At this stage the many-electron wavefunctions were replaced by their FS-RCCSD model space parts, i.~e. by the eigenvectors of the effective Hamiltonians. It is worth noting that the resulting SO-deperturbed states, in a strict analogy with their empirical counterparts, incorporate implicitly the effects of SO coupling with the scalar state outside of the chosen set.

\subsection{Analytical representation of the interatomic potentials and spin-orbit functions}\label{analytic}

The overwhelming number of experimental terms investigated in the present work can be unambiguously assigned to $A^1\Sigma^+_u(0^+_u)$ or $b^3\Pi_u(0^+_u)$ state of the $A\sim b\sim c \sim B$ complex, so it is not surprising that the global CC deperturbation model depends most strongly on interatomic $U_{A}(R)$ and $U_{b0}(R)$ potentials of the interacting $A$ and $b$ states, as well as on the relevant SO-splitting $A_{so}(R)$ and SO-coupling $\xi_{Ab0}(R)$ functions. 

To improve the optimization and extrapolation facilities of the global DDF both $A$ and $b$ state PECs were represented in fully analytical ‘‘Double-Exponential/Long-Range’’ (DELR) form~\cite{LeRoyDELR}:
\begin{eqnarray}\label{DELR}
U^{DELR}(R) = [T_{dis} + (Az - B)z] - U^{LR},
\end{eqnarray}
where coefficients $A$ and $B$ are determined as
\begin{eqnarray}\label{DELRAB}
A &=& D_e - U^{LR}(R_e) - \frac{dU^{LR}/dR|_{R_e}}{\beta(R_e)} ;\\
B &=& D_e - U^{LR}(R_e) + A ,\nonumber
\end{eqnarray}
$D_e=T_{dis}-U^{DELR}(R_e)$ is the well depth of the potential and $T_{dis}$ is the dissociation energy with respect to the minimum of the ground $X$-state. The symbol $R_e$ involved into the $z = e^{-\beta(R-R_e)}$ variable of Eq.(\ref{DELR}) and into Eq.(\ref{DELRAB}) means the equilibrium distance (minimum) of DELR potential while the coefficient $\beta$  
\begin{equation} \label{V_DELR_beta}
\beta(R)=\sum^{N}_{i=0}\beta_iy^i;\quad  y \equiv \frac{(R/R_{ref})^p - 1}{(R/R_{ref})^p + 1}
\end{equation}
is the polynomial function of the reduced radial coordinate $y\in [-1,1]$, where $R_{ref}$ is the reference distance and $p$ is the integer parameter. 

The $T_{dis}$ value in Eq.(\ref{DELR}) relates on the dissociation limit $T_{dis}^b$ of the $b0^+_u$ state and the well depth $D_e^X$ of the ground state
\begin{eqnarray}\label{Diss}
T_{dis}\equiv T_{dis}^A = T_{dis}^b + \xi^K_{so} = D_e^X + \nu^K_{4p-4s},
\end{eqnarray}
through the well known energy~\cite{Tiecke2019} of the K($4p$)-K($4s$) atomic transition $\nu^K_{4p-4s}=13023.6587$ cm$^{-1}$ and the SO-splitting of the K($4p$) state $\xi^K_{so}=[E^K_{4p\;^2{\rm P}_{\frac{3}{2}}}-E^K_{4p\;^2{\rm P}_{\frac{1}{2}}}]/3=19.2368$ cm$^{-1}$.

The long-range part of DELR potential was approximated by the leading term
\begin{eqnarray}\label{LR}
U^{LR}(R) = \frac{C^{\Sigma/\Pi}_3}{R^3}f_{ret}^{\Sigma/\Pi}, 
\end{eqnarray}
which accounts for the retardation effect
\begin{eqnarray}\label{retard}
f_{ret}^{\Sigma}(R)&=&\cos\left(\frac{R}{\lambdabar}\right)+
\left(\frac{R}{\lambdabar}\right) \sin\left(\frac{R}{\lambdabar}\right); \\
f_{ret}^{\Pi}(R)&=&f_{ret}^{\Sigma}(R)-
\left(\frac{R}{\lambdabar}\right)^2 \cos\left(\frac{R}{\lambdabar}\right)\nonumber; 
\quad 2\pi\lambdabar=\frac{c}{\nu^K_{4p-4s}} 
\end{eqnarray}
where the leading dispersion coefficients $C^{\Sigma/\Pi}$ relate on the dipole matrix element of non-relativistic K($4p$)-K($4s$) transition and the radiative lifetime of the K($4p$) state:
\begin{eqnarray}\label{tau_K}
|d^K_{4p-4s}|^2=C^{\Pi}_3=\frac{C^{\Sigma}_3}{2};\qquad
\tau^K_{4p} = \frac{3\hbar\lambdabar}{2C^{\Sigma}_3}.
\end{eqnarray}

Both diagonal $A_{so}$ and off-diagonal $\xi_{Ab0}$ SO functions involved into Eq.(\ref{Ham_diag}) and Eq.(\ref{Ham_offdiag1}), respectively, were represented as the sum of the fixed $\xi^{ab}_{so}$ and variable $\Delta \xi_{so}$ parts:
\begin{eqnarray}\label{DeltaSO}
\xi_{so}(R) = \xi^{ab}_{so}(R) + \Delta \xi_{so}(R).
\end{eqnarray}

The fixed part $\xi^{ab}_{so}$ consists of the relevant \emph{ab initio} point-wise data, which were smoothly interpolated by cubic splines with the appropriate boundary conditions~\cite{NR}, namely: $d^2\xi_{so}/dR^2|_{R_{min}}=d\xi_{so}/dR|_{R_{max}}=0$. 

The variable part $\Delta \xi_{so}$ was defined as the Chebyshev polynomial expansion (CPE) with the correct long-range behaviour at the dissociation limit\cite{Bormotova2019}:
\begin{eqnarray}\label{CPESO}
\Delta \xi_{so}(R)=\frac{\sum_{i=1}^{N}\alpha_iT_{i-1}(y)}
{1+(R/R_{ref})^p},
\end{eqnarray}
where $T_i$ are the 1-st kind Chebyshev polynomials with respect to the reduced variable 
\begin{eqnarray}\label{CPEyp}
y \equiv \frac{(R/R_{ref})^p - 1 }{(R/R_{ref})^p + 1-2(R_{min}/R_{ref})^p} ,
\end{eqnarray}
which transforms the conventional semi-interval $R\in [R_{min},+\infty)$ into the finite domain $y\in [-1,+1]$.

\subsection{The least-squares processing}\label{fit}

The trial parameters of the $U^{DELR}_A(R)$ and $U^{DELR}_{b0}(R)$ analytical potentials (see Eq.(\ref{DELR})) for the $A0^+_u$ and $b0^+_u$ states, as well as the CPE coefficients $\alpha_i$ of both $\Delta \xi_{Ab0}(R)$ and $\Delta A_{so}(R)$ correction functions (see Eqs.(\ref{DeltaSO}) and (\ref{CPESO})) were refined in the framework of the weighted nonlinear least-square fitting (NLSF) procedure:
\begin{eqnarray}\label{chisquared}
\chi^2=\chi^2_{expt} + w\chi^2_{ab} ,
\end{eqnarray}
which simultaneously accounts for the contribution of the experimental term values and the {\it ab initio} data in the object $\chi^2$-function. The \emph{ab initio} data were incorporated into the NLSF procedure in order to maintain a physically correct behaviour of the empirical functions outside the experimentally available region of the internuclear distance. During the iterative minimization of the functional (\ref{chisquared}) the weighting factor $w$ is smoothly varied to provide a proper balance between the contributions of experimental and {\it ab initio} data into to the total $\chi^2$-value.

The first term in the expression (\ref{chisquared}) has been evaluated as the sum 
\begin{eqnarray}\label{chiexpt}
\chi^2_{expt}=\sum_{i=1}^{N_{set}}\left [\sum_{j=1}^{N_{expt}}\left(\frac{E^{expt}_j(J^{\prime})-E^{CC}_j(J^{\prime})+\Delta^{shift}_i}{\sigma^{expt}_i}\right)^2\right],
\end{eqnarray}
where $N_{expt}$ is the total number of rovibronic levels of the particular experimental data set (see Table~\ref{tab:dataset}) included in the NLSF procedure. Only assigned to the most abundant $^{39}$K$_2$ isotopologue term values were used in the fit. $E^{CC}_j(J^{\prime})$ are the non-adiabatic eigenvalues calculated according to the CC equations (\ref{CC}), while $E^{expt}_j(J^{\prime})$ are the experimental term values with their uncertainties $\sigma^{expt}_i$. Here, $\Delta^{shift}_i$ is the systematic energy shift presumably attributed to the tacit uncertainty of the experimental term values corresponding to the $i$-th data set.

The last term in Eq.(\ref{chisquared}) was estimated as the double sum:  
\begin{eqnarray}\label{chiab}
\chi^2_{ab} = \sum_{i\in A0^+_u,b0^+_u}\left [\sum_{j=1}^{N_{ab}}\left (\frac{U^{DELR}_i(R_j) - U^{ab}_i(R_j)}{\sigma^{ab}_j}\right)^2\right ],
\end{eqnarray}
where $N_{ab}$ is the total number of the internuclear distance points $R_j$ where the relevant {\it ab initio} point-wise potentials $U^{ab}_i(R)$ were evaluated within the framework of scalar-relativistic calculations (see Sec.~\ref{abinitio}). The uncertainties $\sigma^{ab}_j$ were estimated by a comparison of the results of scalar-relativistic and fully relativistic calculations.

\subsection{Numerical recipes} 
 
 The main computational challenge of the DDF procedure is the multiple numerical solution of the six CC radial equations in the entire bound spectral range, starting from the practically isolated triplet levels in a vicinity of the bottom of the $b0^+_u$-state located near 9885 cm$^{-1}$ (with respect to the minimum of the ground $X$-state) and ending with fully mixed $A0^+_u\sim b0^+_u$ levels closely-lying to the $4s^2$S$_{\frac{1}{2}}$+$4p^2$P$_{\frac{1}{2}}$ dissociation threshold with energy 17436.075 cm$^{-1}$. It should be also noted that due to the smooth long-range asymptotic behaviour of interatomic potentials $U^{LR}(R\to +\infty)\approx T_{dis}-C^{\Sigma/\Pi}_3/R^3$ the highly-oscillating multi-component wavefunctions for the highest vibrational levels are distributed in a very wide range of $R$. For instance, the classical left $R_{-}$ and right $R_{+}$ turning points for the last observed $v_A$=245 level~\cite{Lisdat2006, Falke2007PhD} of the $A\sim b$ complex with the bound energy only about -0.0164 cm$^{-1}$ seems to be located at $R_{-}\approx$ 2.6~\AA~and $R_{+}\geq$ 300~\AA, respectively.   

To improve the efficiency of the iterative solution of the CC equations we implemented the analytical mapping procedure~\cite{Meshkov2008} based on replacement of the conventional radial coordinate $R$ by the reduced counterpart
\begin{eqnarray}\label{ycoordinate}
y \equiv \frac{(R/R_{ref})^p-1}{(R/R_{ref})^p+1};\qquad y\in[-1,+1],
\end{eqnarray}
where the optimal parameters of mapping $R_{ref}$=5.0~\AA~and $p$=1.5 were adopted to minimize a truncation error in the merged eigenvalues evaluated in the entire energy range of the complex treated. The analytically (exactly) transformed CC equations were then solved on the finite interval $R\in [2.3,10~000]$~\AA~by the central five-points finite-difference (5FD) boundary value method, which enforced the fixed number $M$ of the uniformly distributed $y$-points. The integral form of the error correction to eigenvalues appeared due to the 5FD method~\cite{Meshkov2008} was applied for extrapolating the calculated energies to infinite number of grid points $E^{CC}_j(M\to +\infty)$. Thus, the mapping procedure had reduced the overall absolute error in all calculated energies below 0.0005 cm$^{-1}$ by using only $M$=10~000 uniformed grid $y$-points.

The particular eigenvalue and eigenfunction problem of the resulting band matrix was solved by the implicitly restarted Lanczos
method in the shift-inverted spectral transformation mode implemented in ARPACK package~\cite{ARPACK}. The minimum of the functional (\ref{chisquared}) was searched by the modified Levenberg-Marquardt algorithm realized by MINPACK software~\cite{MINPACK}. A quantum phase formalism~\cite{ABARENOV} was applied for solving the single-channel problem corresponding to the isolated ground $X$-state.

\subsection{Simulation of intensity distributions in the $A\sim b\to X$ LIF progressions}\label{RID}

Since DDF was accomplished according to Sec.~\ref{DDF} by exploiting exclusively the energy data $E^{CC}_j(J^{\prime})$ of the single $^{39}$K$_2$ isotopologue, the reliability of the corresponding multi-component vibrational eigenfunctions $\Phi^{CC}_j(R)$ simultaneously obtained from the solution of CC equations (\ref{CC}) can be independently validated by comparing the relative intensity distributions in the $A\sim b\to X$ LIF spectra with their experimental counterparts.

The line intensity $I^{calc}_{jX}(v_X)$ in the $A\sim b\to X(v_X)$ LIF progressions of $^{39}$K$_2$ and $^{39}$K$^{41}$K isotopologues from the rovibronic $j$-level of the $A\sim b\sim B \sim c$ complex to rovibronic levels of the isolated $X^1\Sigma^{+}_{g}$ state can be evaluated as
\begin{eqnarray}\label{Itensinglet}
I^{calc}_{jX} \sim \nu^4_{jX}\left |M_{jX}\right |^2&\approx& \nu^4_{jX}|\langle \phi_A^{J^{\prime}}|d^{ab}_{AX}|\phi_X^{J^{\prime}\pm 1}\rangle|^2; \\
\nu_{jX} = E_j^{CC}(J^{\prime})-E_X^{J^{\prime} \pm 1}&;&\quad 
M_{jX} = \sum_{i=1}^6 
\langle \phi_i^{J^{\prime}}|d_{iX}|\phi_X^{J^{\prime}\pm1} \rangle, \nonumber
\end{eqnarray}
where the contribution of both spin-forbidden $b-X$, $c-X$ and spin-allowed $B-X$ transitions was completely neglected. The latter assumption is justified by the extremely small admixture of the $B$ state for all levels of the complex selected for the intensity testing, see Fig.~\ref{Fig3}. Thus, $\phi_A^{J^{\prime}}(R)$ in the right side of Eq.(\ref{Itensinglet}) is the singlet $A$ state fraction of the multi-component wavefunction corresponding to the eigenvalue $E_j^{CC}(J^{\prime})$, whereas $d^{ab}_{AX}(R)$ is the spin-allowed $A-X$ transition dipole moment evaluated in Ref.\cite{Krumins2020} using the robust {\it ab initio} FS-RCC finite-field (FS-RCCSD-FF) method~\cite{ZaitsevskiiTDM}. The adiabatic energies $E_X^{J^{\prime} \pm 1}$ and eigenfunctions $\phi_X^{J^{\prime}\pm 1}(R)$ for the ground state were obtained by solving the radial equation with empirical $X$-state PEC\cite{Tiemann2020}.

\section{Results and discussion}


\begin{figure}
\includegraphics[scale=0.35]{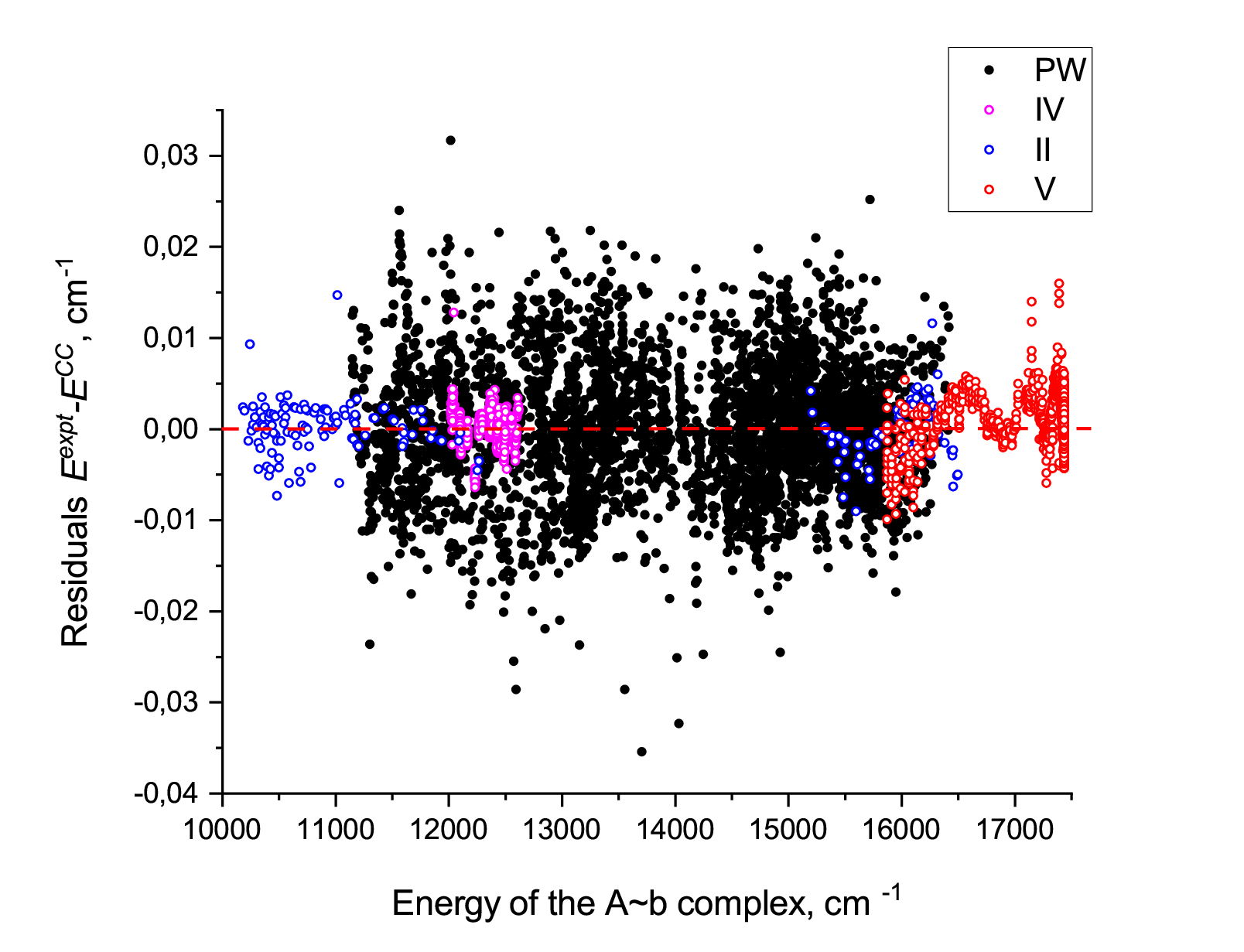}
\caption{Residuals $E^{expt}_j-E^{CC}_j$ between the experimental and fitted term values as dependent on the $A\sim b$ complex energy for $^{39}$K$_2$: black points - present (PW), magenta ~\cite{Lisdat2001, Lisdat2001PhD}, red ~\cite{Lisdat2006, Falke2007PhD}, blue ~\cite{Manaa2002}. The legend shows relation to the data sets in Table~\ref{tab:dataset}.}
\label{residuals}
\end{figure}


\begin{figure}
\includegraphics[scale=0.4]{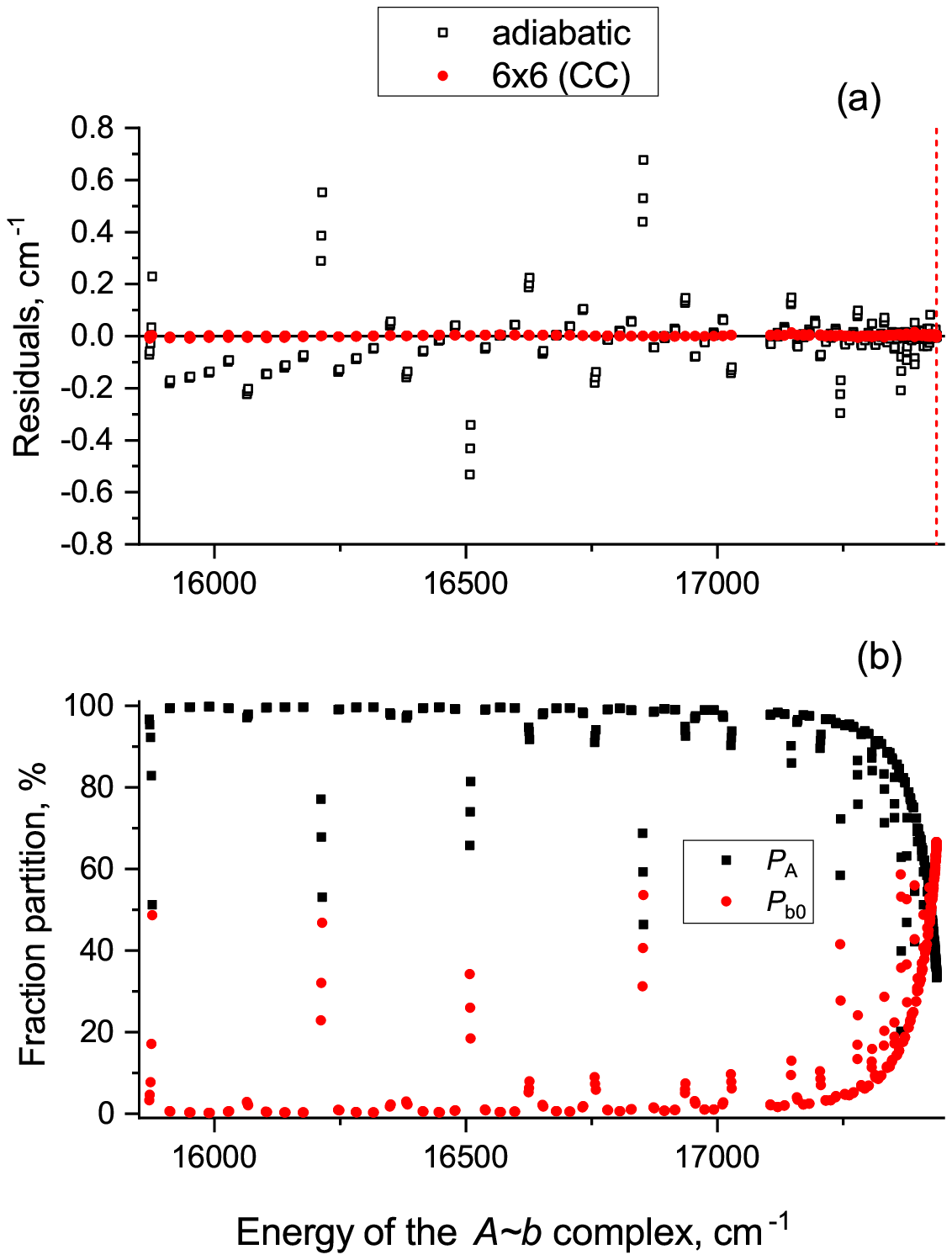}
\caption{ (a) Residuals $E^{expt}_j-E^{CC}_j$ between the highly-excited term values of the $^{39}$K$_2$ $A\sim b$ complex precisely measured by the 2s-LES-MB setup~\cite{Falke2007PhD} and their theoretical counterparts calculated in~\cite{Lisdat2006} under the simple adiabatic approximation (the isolated SO-deperturbed $A$ state) and presently by using the advanced non-adiabatic 6$\times$6 CC approach; vertical dashed line denotes the dissociation threshold  with energy 17436.075 cm$^{-1}$; (b) the fraction partition $P_A$, $P_{b0}$ of the rovibronic levels of the $A\sim b$ complex with the vibrational quantum numbers $v_A\in[88,245]$ and $v_{b0}\in[88,236]$. These $v_A$ and $v_{b0}$ values have been determined as a number of nodes of the corresponding non-adiabatic vibrational wavefunctions $\phi_A(R)$ and $\phi_{b0}(R)$, respectively.}
\label{residualsFalke}
\end{figure}

\subsection{Experimental and calculated rovibronic term values of the $A\sim b$ complex}

The K$_2$ rovibronic term values data considered in present study are summarized in Table~\ref{tab:dataset}; all data sets from different sources can be find in Supplementary Material~\cite{SM}. Only the term values data for the most abundant $^{39}$K$_2$ isotopologue from the sets I to V in Table~\ref{tab:dataset} were included in the fit, see Fig.~\ref{Fig3}a; they contain more than 5700 term values, from which 4053 values (I) are obtained in the present work (PW). The compiled data set covers the whole energy range from 9955 cm$^{-1}$ to 17436 cm$^{-1}$, that is, from the bottom of the $b^3\Pi_u(b0^+_u)$ state (data III from Ref.~\cite{Kim95}) up to the dissociation limit (data V from Ref.~\cite{Falke2007PhD}). The term values of highly excited levels were derived by us for both main and mixed istopologues by adding the original sub-Doppler line positions measured in Ref.~\cite{ Falke2007PhD} to the ground $X$-state levels calculated using the highly accurate adiabatic empirical PEC from Ref.~\cite{Tiemann2020}. These values are found to be close within 0.003-0.005 cm$^{-1}$ to their previous Dunham-based counterparts~\cite{Lisdat2006, Falke2007PhD} corrected for the $Y^X_{00}$-value~\cite{Amiot1995}.

It should be noted that, as mentioned in Sec.~\ref{fit}, the fitting routine, which exploited data sets from different sources, contained the systematic energy shift $\Delta_i^{shift}$, see Eq.(\ref{chiexpt}). The term values of sets I, IV, and V were fixed at the DDF procedure ($\Delta_i^{shift}\equiv 0$). For two other sets II and III the systematic $\Delta_i^{shift}$-value was applied as described in Sec.~\ref{fit}. The obtained correction $\Delta_i^{shift}$ = +0.031 cm$^{-1}$ has been adjusted to fit the PF-OODR data from Ref.~\cite{Manaa2002}. Though the reason for this shift is not clear, it is worth to mention that the calculated $E^{expt}_j$ value for the level $J^{\prime}$ = 66 excited in the first step of the PF-OODR experiment is perfectly reproduced, see Sec.~\ref{not_incl} below.

The most significant correction of -2.1 cm$^{-1}$ applied for set III can be at least partially attributed  to the large experimental uncertainty. The reason of including the data from~\cite{Kim95} is to cover the energy range close to the very bottom of the "dark" $b^3\Pi_u$ state with $v_{b0}$ = 0, as well as to provide the experimental data for all three fine $b0^+_u$, $b1^+_u$ and $b2^+_u$ components of the triplet state.  

\begin{table*}
\caption{Summary of the experimental term values of the $A\sim b$ complex of K$_2$ considered in the present work (PW). Symbol $^{\dag}$ marks the sets of the $E_j^{expt}$ and $\Delta_i^{shift}$ values used in Eq.(\ref{chiexpt}) of the DDF procedure. The columns $Mean$ and $SD$ (standard deviation) provide an appropriate statistic of the residual data: $E^{expt}_j-E^{CC}_j+\Delta^{shift}_i$. All energies are in cm$^{-1}$.}
\label{tab:dataset}
\begin{ruledtabular}
\begin{tabular}{ccccccccccc}
N$_{set}$ & isotope & N$_{expt}$ & $J^{\prime}$-values & $E_j^{expt}$-region & Method & Ref. & $\sigma_i^{expt}$ & $\Delta_i^{shift}$ & $Mean$ & $SD$\\ 
\hline
 I & $^{39}$K$_2$ & 4054$^{\dag}$& [2,149] & [11145,16651] & FT-LIF & PW  & 0.015 & - & 0.000 & 0.007\\
 & $^{39}$K$^{41}$K & 705 & [1,162] & [11155,15863] &  &  &  & - & 0.003 & 0.009 \\ 
 & $^{41}$K$_2$ & 14 & [12,86] & [11637,12385] &  &  &  & - & -0.007 &  0.015\\
\hline
 II & $^{39}$K$_2$ & 188$^{\dag}$ & [60,74] & [10180,16497] & PF-OODR & \cite{Manaa2002}  & 0.02& +0.031$^{\dag}$ & 0.000 & 0.003\\ 
 III &   & 17$^{\dag}$  & [20,22] & [9955,10433] &  PF-OODR & \cite{Kim95} & 0.50 & -2.1$^{\dag}$& 0.002 & 0.323 \\
 \hline
 IV & $^{39}$K$_2$ & 511$^{\dag}$ & [0,37] & [12019,12616] & 2s-LES-MB &  \cite{Lisdat2001PhD, Lisdat2001, Manaa2002}  & 0.005 & - & 0.000 & 0.002 \\  
  & $^{39}$K$^{41}$K  & 62  & [0,10] & [12408,12849] &  & \cite{Lisdat2001PhD, Falke2007PhD} & 0.005 & - & 0.000 & 0.001 \\
  \hline
 V & $^{39}$K$_2$  & 1023$^{\dag}$   & [2,16] & [15870,17436] &  2s-LES-MB & \cite{Falke2007PhD} & 0.001-0.005 & - & 0.001 & 0.003\\
   & $^{39}$K$^{41}$K & 29 & [8,10] & [17435.37,17436.06] &  & \cite{Falke2007} & 0.001 & - & 0.000 & 0.001\\
 \hline 
VI & $^{39}$K$_2$ & 327 & [3,177] & [11857,16037] & FT-LIF & \cite{AmiotJMS1991, Amiot1995}  & 0.010 & - & 0.008 & 0.008\\
VII &  & 308 & [4,64] & [11146,12064] & AOTR & \cite{JongJMS1992}& 0.015 & - & 0.005 & 0.008 \\
VIII &  & 315 & [4,96] & [11975,12655] & FT-LIF & \cite{Ross87} & 0.013 &  & -0.001 & 0.009 \\
IX &  & 64 & [12,34] & [12557,13765] & AOTR & \cite{Lyyra1990,JongJMS1992} & 0.015 & - & 0.006 & 0.012 \\
X &  & 65 & [6,52] & [11307,12066] & AOTR & \cite{Manaa2002} & 0.017 & - & 0.004  & 0.016 \\
XI &  & 6 & [18,20] & [14889,14990] & AOTR & \cite{Manaa2002} & 0.017 & - & -0.01 & 0.03 \\
\end{tabular}
\end{ruledtabular}
 FT-LIF  - Fourier-transform of the laser-induced fluorescence spectra; 
 PF-OODR - perturbation facilitated optical-optical double resonance spectroscopy; 
 AOTR - all-optical triple-resonance spectroscopy; 
2s-LES-MB - two steps laser excitation spectroscopy of a molecule beam
\end{table*}

\begin{table}
\caption{The resulting parameters of the analytical DELR potentials obtained for the deperturbed singlet $A^1\Sigma^+_u(0_u^+)$ and triplet $b^3\Pi_u(0_u^+)$ states of K$_2$. The dispersion coefficient $C^{\Sigma}_3$ is defined in cm$^{-1}$/\AA$^3$, internuclear distances in~\AA, and energies in cm$^{-1}$. The polynomial coefficients $\beta_i$ are dimensionless and their total number for both states is $N$=17. The parameters with more digits are presented in Supplementary Material~\cite{SM}.} \label{DELRpar}
\begin{tabular}{ccc}
\hline \hline
Parameter & $b^3\Pi_u(0_u^+)$-state & $A^1\Sigma^+_u(0_u^+)$-state \\
\hline
$T_{dis}$ & \multicolumn{2}{c}{17474.5688}\\
$C^{\Sigma}_3=2C^{\Pi}_3$ & \multicolumn{2}{c}{550100}\\
$R_{\textrm{ref}}$ & \multicolumn{2}{c}{5.0}\\
$p$ & \multicolumn{2}{c}{3}\\
\hline
 ${\mathfrak D}_e$  & 7570.4715 & 6367.5309\\   
 $R_{e}$ & 3.888657 & 4.550638\\
\hline
$\beta_{0}$  &  0.68284220  &  0.54006196 \\
$\beta_{1}$  &  0.20408495  &  0.06155514 \\
$\beta_{2}$  &  0.19374658  &  0.09689141 \\
$\beta_{3}$  &  0.18735892  &  0.08927925 \\
$\beta_{4}$  &  0.09070268  &  0.10889245 \\
$\beta_{5}$  & -0.02173617  &  0.36613441 \\
$\beta_{6}$  &  0.00039517  & -0.00000448 \\
$\beta_{7}$  & -0.73601292  & -2.03437500 \\
$\beta_{8}$  & -2.86990662  &  0.10673692 \\
$\beta_{9}$  &  0.91874942  &  9.23970194 \\
$\beta_{10}$ &  9.57662145  &  0.00595872 \\
$\beta_{11}$ & -1.32818589  & -23.0068544 \\
$\beta_{12}$ & -17.1303243  &  1.65936885 \\
$\beta_{13}$ &  0.84419525  &  30.1200697 \\
$\beta_{14}$ &  16.3070732  & -7.48445214 \\
$\beta_{15}$ & -0.20286255  & -18.2492960 \\
$\beta_{16}$ & -6.27975991  &  7.38151726 \\
$\beta_{17}$ &  0.00237189  &  1.52291107 \\
\hline
 $T_{e}$ &  9884.8605 & 11107.0378 \\  
\hline
\end{tabular}
\end{table}

\begin{table}
\caption{The resulting (both fitted and constrained) parameters obtained for the variable parts of diagonal $\Delta A_{so}$ and off-diagonal $\Delta \xi_{Ab0}$ SO functions of K$_2$.
The fitting coefficients $\alpha_i$ are in cm$^{-1}$, the fixed parameters $R_{\rm{min}}$ and $R_{\rm{ref}}$ are in~\AA. The fitting parameters $\eta_{Ab1}$ and $\eta_{bb}$ involved in Eq.(\ref{Ham_offdiag1}) are dimensionless. The parameters with more digits are presented in Supplementary Material~\cite{SM}.}\label{SOF}
\begin{tabular}{ccc}
\hline \hline  
& $\Delta \xi_{Ab0}$ & $\Delta A_{so}$\\
\hline 
$p$ & \multicolumn{2}{c}{3}\\
$R_{\rm{min}}$ & \multicolumn{2}{c}{1.99}\\
$R_{\rm{ref}}$ & 4.0 & 5.0\\
\hline
$\alpha_1$ & 0.19985938 &  4.13434831 \\
$\alpha_2$ & 2.11715093 & -4.10868122 \\
$\alpha_3$ & 0.00396364 &  2.90669862 \\
$\alpha_4$ & 2.34082411 & -2.55234792 \\
$\alpha_5$ & 0.36215827 &  0.75114018 \\
\hline
 &$\eta_{Ab1}$ & $\eta_{bb}$\\
\hline
 & -0.005175 & 0.006823\\
\hline
\end{tabular}
\end{table}

\begin{table}
\caption{Empirical and {\it ab initio} adiabatic electronic energies $T_e$ and equilibrium distances $R_e$ of the $A^1\Sigma^+_u(A0_u^+)$ and $b^3\Pi_u (b0_u^+, b1_u, b2_u)$ states of K$_2$. Here SO-dep means SO-deperturbed method. Symbol $^{\dag}$ corresponds to the SO-deperturbed FS-RCC results.} \label{TeRe}
\begin{tabular}{cllll}
\hline\hline
State  & $T_e$(cm$^{-1}$) & $R_e$(\AA) & Method & Ref.\\
\hline
$A^1\Sigma^+_u$ & 11107.038 & 4.55064 & 6$\times$6 CC & PW \\
                & 11107.019 & 4.551 & SO-dep & \cite{Lisdat2006}\\
                & 11107.112 & 4.54658 & 4$\times$4 CC & \cite{Manaa2002}\\
                & 11107.141 & 4.55078 & 4$\times$4 CC & \cite{Lisdat2001}\\
                & 11107.92 & 4.5498 & EHA & \cite{JongJMS1992}\\
                & 11105.9 & 4.557 & EHA & \cite{Ross87}\\[0.5ex]
                & 11041  & 4.53 & CI-CPP-$l$ & \cite{Magnier2004}\\ 
                & 11256  & 4.57 & CI-CPP  & PW \\
                & 11053 & 4.555 & FS-RCC$^{\dag}$ & PW \\[1.5ex]  
$A0_u^+$        & 11050 & 4.561 & FS-RCC & PW \\  
                & 11045 & 4.55 & CI-CPP-$l$ & \cite{Magnier2009}\\
\hline
$b^3\Pi_u(0_u^+)$ & 9884.861 & 3.8887 & 6$\times$6 CC & PW \\[0.5ex]
                  & 9903 & 3.891 & FS-RCC$^{\dag}$ & PW \\[1.5ex]
$b0_u^+$          & 9903 & 3.891 & FS-RCC & PW \\
                  & 9805 & 3.86 & CI-CPP-$l$ & \cite{Magnier2009}\\
\hline
$b^3\Pi_u(1_u)$ & 9907.20 & 3.8905 & 6$\times$6 CC & PW \\
                  & 9907.50 & 3.8437 & 4$\times$4 CC & \cite{Manaa2002}\\
                  & 9912.95 & 3.8945 & 4$\times$4 CC & \cite{Lisdat2001}\\
                  & 9911.8 & 3.875 & EHA & \cite{JongJMS1992}\\
                  & 9910.0 & 3.883 & EHA & \cite{Ross87}\\[0.5ex]
                  & 9827 & 3.88 & CI-CPP-$l$ & \cite{Magnier2004}\\ 
                  & 9993 & 3.91 & CI-CPP  & PW \\[1.5ex]            
$b1_u$            & 9925 & 3.892 & FS-RCC & PW \\ 
                  & 9832 & 3.86 & CI-CPP-$l$ & \cite{Magnier2009}\\
\hline
$b^3\Pi_u(2_u)$ & 9929.531 & 3.8931 & 6$\times$6 CC & PW \\[1.5ex]
$b2_u$            & 9948 & 3.893 & FS-RCC & PW \\
                  & 9859 & 3.86 & CI-CPP-$l$ & \cite{Magnier2009}\\
\hline
\end{tabular}
\end{table}

The data for presently measured isotopologues $^{39}$K$^{41}$K and $^{41}$K$_2$ have been used for verifying the quality of the present fit, see Sec.~\ref{ISE}. The data for $^{39}$K$_2$ isotopologue not included in the fit, see sets from VI to XI in Table~\ref{tab:dataset}, belong to the energy range, which is well covered by the present experiment. Therefore these data have been used for verifying the prediction capability of the present fit, see Sec.~\ref{not_incl}. All term values of data sets summarized in Table~\ref{tab:dataset} can be found in Supplementary Material~\cite{SM}. 

The residuals $E_j^{expt} - E_j^{CC}$  between the experimental and fitted term values are presented in Fig.~\ref{residuals}. As is seen, most of residuals (ca. 95 \%) lay within $\pm$0.01 cm$^{-1}$ interval, which corresponds to SD about 0.007 cm$^{-1}$ (see the SD values in Table~\ref{tab:dataset}). It should be noted that the data (set III) from ~\cite{Kim95} are not depicted in the figure since the residuals would be markedly out of scale. It is important that no systematic shift (see the $Mean$ value) of residuals exploiting the fixed values of data sets from different sources (PW, IV and V, see Table~\ref{tab:dataset}) has been observed.

To characterize the particular improvement of description of the highly-excited rovibronic K$_2$  $A\sim b$ levels by the current 6$\times$6 CC DDF procedure, we present in Fig.~\ref{residualsFalke}a the residuals between the term values produced by the present fit and the experimental values given in~\cite{Falke2007PhD}. For a comparison the residuals between the same experimental values~\cite{Falke2007PhD} and the adiabatic model predictions obtained in the framework of the isolated SO-deperturbed $A$ state approximation~\cite{Falke2007}, is also depicted. 

As can be seen, the present deperturbation model describes all data over 16 000 cm$^{-1}$, including the levels under strong local perturbations, with $SD$ about 0.003 and $Mean$ value of only 0.001 cm$^{-1}$, see Fig.~\ref{residuals} and Table~\ref{tab:dataset}. The positions of local perturbations are clearly exposed in the respective dependencies of  fraction partitions $P_A$ and $P_{b0}$, see Fig.~\ref{residualsFalke}b. As expected, in a vicinity of the dissociation threshold the mixing between $A$ and $b0^+$ states steeply increases as the energy of the $A\sim b$ complex increases, and the fraction partitions $P_A$ and $P_{b0}$ monotonically approach their asymptotic values of $\frac{100}{3}\approx 33.3$\% and $\frac{200}{3}\approx 66.6$\%, respectively. 

\begin{table}
\caption{Comparison of experimental dissociation energies $D_e^X$ (cm$^{-1}$) available for the ground $X$-state of $^{39}$K$_2$.} \label{tab:Dx}
\begin{ruledtabular}
\begin{tabular}{ccccc}
 PW & \cite{Falke2006} & \cite{Asen2008} & \cite{Tiemann2020} & \cite{Falke2008}\\
\hline
4450.910(5) & 4450.904(4) & 4450.906(50) & 4450.906(3) & 4450.899(4) \\
\end{tabular}
\end{ruledtabular}
\end{table}

\begin{table}
\caption{Comparison of long-range coefficients $C_3^{\Sigma}$ ($\times10^5$ cm$^{-1}$\AA$^3$) related to the K($4s$)+K($4p$) asymptotes and radiative lifetimes $\tau^K$ (ns) of the $4p\;^2{\rm P}_{\frac{1}{2}}$ and $4p\;^2{\rm P}_{\frac{3}{2}}$ fine components of the K atom.} \label{tab:C3tau}
\begin{ruledtabular}
\begin{tabular}{cccc}
$C_3^{\Sigma}$ & $\tau^K_{4p\;^2{\rm P}_{\frac{1}{2}}}$ & $\tau^K_{4p\;^2{\rm P}_{\frac{3}{2}}}$ & Ref.\\
\hline
\multicolumn{4}{c}{Empirical}\\
5.501(4) & 26.65(3) & 26.30(3) & PW \\
5.483(6) & 26.74(3) & 26.39(3) & \cite{Falke2006} \\
5.493(9) & 26.69(5) & 26.34(5) & \cite{WangPRA97} \\ 
5.487(9) & 26.72(5) & 26.37(5) & \cite{Wang97}\\
\multicolumn{4}{c}{{\it Ab initio}}\\
5.536  & 26.52 & 26.17 & \cite{ZaitsevskiiTDM, Krumins2020}\\
\end{tabular}
\end{ruledtabular}
\end{table}

\begin{table*}
\caption{Comparison of some experimental rovibronic term values $E_j^{expt}$ obtained for the $^{39}$K$_2$ $A\sim b$ complex with their theoretical counterparts $E_j^{CC}$ predicted by using the present 6$\times$6 CC deperturbation model. All energies $E^{expt}_j$ presented here include the $Y^X_{00}$~ =~${-}$0.022 cm$^{-1}$ correction. The energies are given in cm$^{-1}$ and the fraction partition in \%.}
\label{tab:termtest}
\begin{ruledtabular}
\begin{tabular}{lrllrrrrr}
Ref. &	$J^{\prime}$ & $E_j^{expt}$ & $E_j^{CC}$ & $E_j^{expt}-E_j^{CC}$ & $P_A$ & $P_{b0}$ & $P_{b1}$ & $P_{b2}$ \\
\hline
\cite{Chu2005} 
&24 & 11718.182 & 11718.197 &-0.015 & 35.0 & 64.6 & 0.4 & 0.0\\
&26 & 11726.741 & 11726.750 &-0.009 & 48.0 & 51.5 & 0.5	& 0.0\\
&28 & 11732.098	& 11732.088 & 0.010 & 30.2 & 69.0 & 0.8 & 0.0\\
&30 & 11738.064	& 11738.054 & 0.010 & 17.7 & 81.3 & 1.0 & 0.0\\
&32	& 11744.578	& 11744.582 &-0.004	& 10.6 & 88.2 & 1.2 & 0.0\\
&34	& 11751.607 & 11751.617 &-0.010	&  6.8 & 91.8 & 1.4 & 0.0\\
&36	& 11612.841	& 11612.808	& 0.033 &  8.0 &  0.7 &89.8	& 1.5\\
&36	& 11759.110	& 11759.123 &-0.013	&  4.7 & 93.8 & 1.5 & 0.0\\
&38	& 11621.363	& 11621.348 & 0.015 & 24.3 &  3.7 &70.7 & 1.3\\
&40	& 11629.743	& 11629.714 & 0.029 &  3.3 &  2.7 &92.1 & 1.9\\
&22	& 12044.480	& 12044.476 & 0.004 & 18.3 & 81.3 & 0.4 & 0.0\\
&22	& 11985.501	& 11985.482 & 0.019 & 19.1 &  0.1 &80.3 & 0.5\\
\cite{Manaa2002} &66 &12415.216	&12415.220& -0.004 & 5.6 & 89.6	& 4.7 & 0.1\\
\cite{Sherstov2007} &25 &12297.748	&12297.743& 0.005 & 1.9 & 97.4 & 0.7 & 0.0 \\
&26 &12300.464	&12300.460&	0.004 & 2.0 & 97.2 & 0.8 & 0.0 \\
\cite{Asen2008} &24	&11718.210	&11718.197	&0.013	&35.0	&64.5	&0.5	&0.0\\
&26	&11726.761	&11726.750	&0.011	&48.0	&51.5	&0.5	&0.0\\
&28	&11732.101	&11732.088	&0.013	&30.2	&69.0	&0.7	&0.0\\
&30	&11738.059	&11738.054	&0.005	&17.7	&81.3	&1.0	&0.0\\
&32	&11744.591	&11744.582	&0.009	&10.6	&88.2	&1.2	&0.0\\
&26	&11726.760	&11726.750	&0.010	&48.0	&51.5	&0.5	&0.0\\
&24	&11718.211	&11718.197	&0.014	&35.0	&64.5	&0.4	&0.0\\
&26	&11726.761	&11726.750	&0.011	&48.0	&51.5	&0.5	&0.0\\
&32	&11744.593	&11744.582	&0.011	&10.6	&88.2	&1.2	&0.0\\
&24	&11718.211	&11718.197	&0.014	&35.0	&64.6	&0.4	&0.0\\
&28	&11732.101	&11732.088	&0.013	&30.2	&69.0	&0.8	&0.0\\	
 \cite{Falke2008}
&6	&17261.487	&17261.486	&0.001	&95.5	&4.5	&0.0	&0.0\\
&8	&17261.787	&17261.786	&0.001	&95.5	&4.5	&0.0	&0.0\\
&10	&17262.167	&17262.166	&0.001	&95.5	&4.5	&0.0	&0.0\\
&2	&17292.516	&17292.516	&0.000	&93.7	&6.3	&0.0	&0.0\\
&6	&17292.839	&17292.839	&0.000	&93.7	&6.3	&0.0	&0.0\\
&10	&17293.449	&17293.449	&0.000	&93.8	&6.2	&0.0	&0.0\\
&10	&17361.247	&17361.242	&0.005	&84.5	&15.5	&0.0	&0.0\\
&6	&17360.821	&17360.816	&0.005	&84.6	&15.4	&0.0	&0.0\\
\end{tabular}
\end{ruledtabular}
\end{table*}

\subsection{{\it Ab initio} and empirical interatomic potentials and non-adiabatic coupling functions of the $A\sim b\sim c\sim B$ complex}


\begin{figure}[t!]
\includegraphics[scale=0.4]{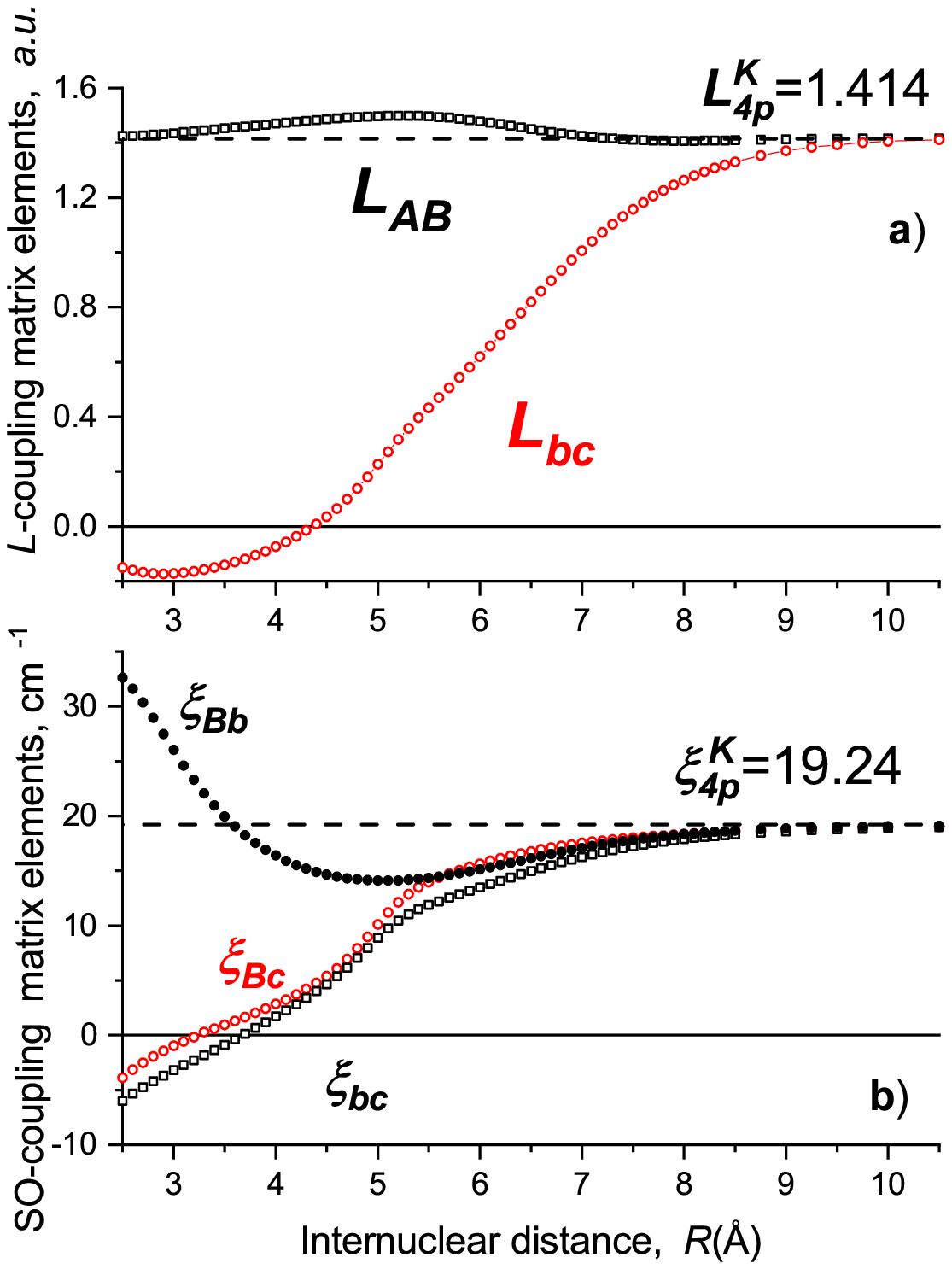}
\caption{The {\it ab initio} $L$-uncoupling (a) and SO coupling (b) matrix elements between the electronic states of the $A\sim b\sim c\sim B$ complex, which were derived within the framework of the CI-CPP approach and were kept in the fixed form during the present DDF. At large internuclear distance $L$-uncoupling functions asymptotically converge to the $\sqrt{l(l+1)}$-value (where $l=1$ for $p$ electron), while the SO matrix elements converge towards the asymptotic $\xi^K_{4p}$-value corresponding to the atomic SO constant of the K($4p^2$P) state. The $L$-uncoupling matrix element between the singlet $A$ and $B$ states obey the Van Vleck's hypothesis\cite{Field2004book} of "pure precession": $L_{AB}(R)\approx \sqrt{L(L+1)-\Lambda(\Lambda\pm 1)}=\sqrt{2}$. } \label{figSOab}
\end{figure}


\begin{figure}[t!]
\includegraphics[scale=0.45]{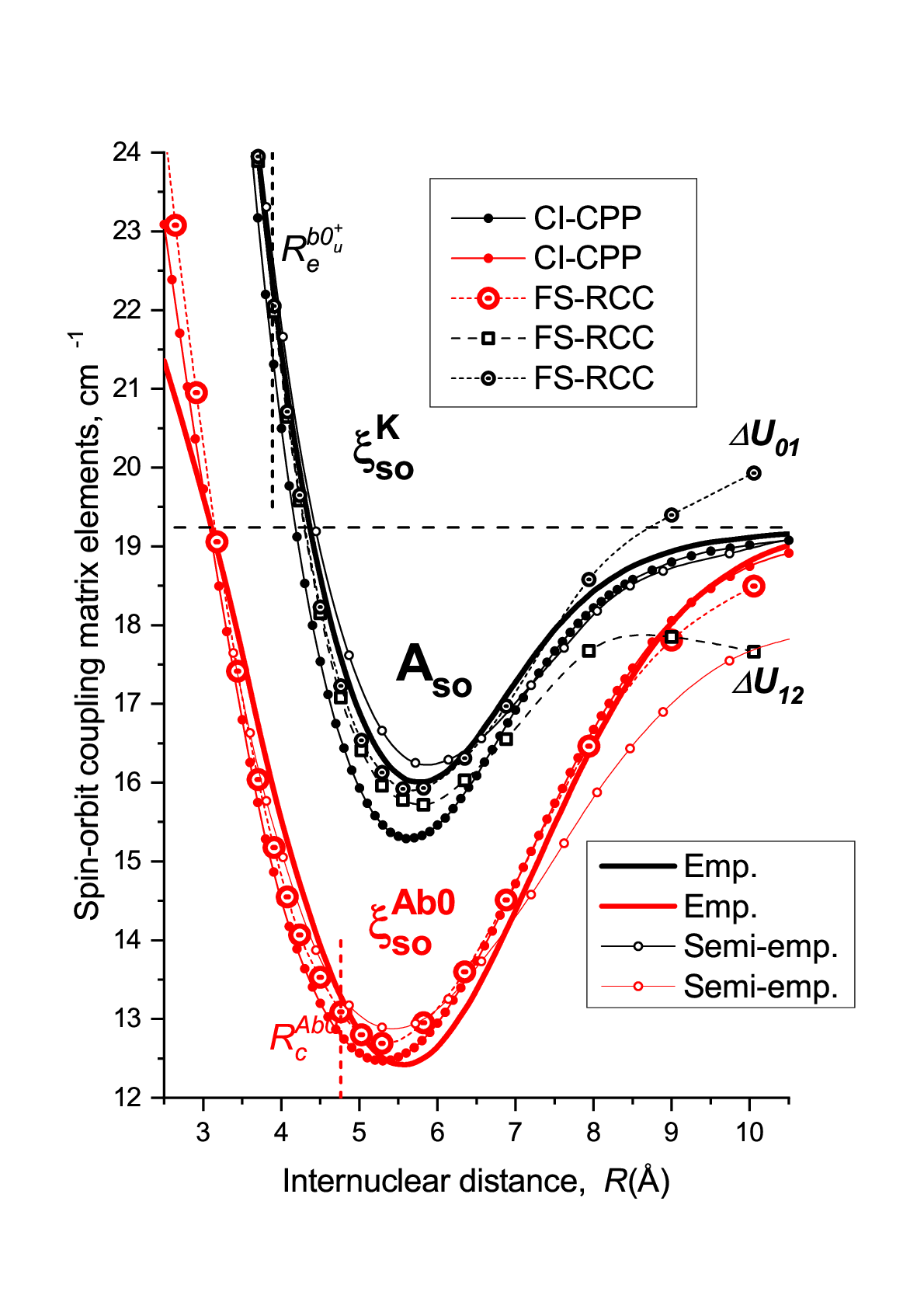}
\caption{Comparison of diagonal $A_{so}$ and off-diagonal $\xi_{Ab0}$ SO functions of K$_2$ obtained within the framework of {\it ab initio} CI-CPP and FS-RCC approaches, respectively, and described in Secs.~\ref{CICPP} and~\ref{FSRCC} with their semi-empirical ~\cite{Manaa2002} and empirical counterparts. $R_e^{b0^+_u}\approx 3.89$~\AA~ is the equilibrium distance of the $b^3\Pi_u(0_u^+)$ state, $R_c^{Ab0}\approx 4.76$~\AA~ is the crossing point of the $A^1\Sigma^+_u$ and $b^3\Pi_u$  Hund's "{\bf{a}}" coupling case PECs, and $\xi^K_{4p}$=19.24 cm$^{-1}$ is the asymptotic value corresponding to the atomic SO constant of K($4p^2$P) state. $\Delta U_{01}\equiv U_{b1_u}-U_{b0^+_u}$ and $\Delta U_{12}\equiv U_{b2_u}-U_{b1_u}$ are the differences of the  Hund's "{\bf{c}}" coupling case PECs obtained for $b0_u^+$, $b1_u$, and $b2_u$ components of the triplet $b$ state during the present fully-relativistic FS-RCCSD calculations.}\label{figSOemp}
\end{figure}
 
The interatomic potentials for all singlet and triplet  $u/g$ states corresponding to the pure Hund's "{\bf{a}}" coupling case and converging to the first four non-relativistic dissociation limits of K$_2$ dimer (see, for instance, Fig.\ref{pec}) have been obtained on the density grid in the range of $R\in [2.3,25]$~\AA~within the framework of the scalar-relativistic CI-CPP electronic structure calculations, which was described in Sec.~\ref{CICPP}. The relevant electronic wavefunctions $\Psi^{sr}_i(\textbf{r};R)$ were then applied for the evaluating of all non-vanishing SO $\xi^{sr}_{ij}(R) = \langle \Psi^{sr}_i|\hat{h^{so}}|\Psi^{sr}_j\rangle_\textbf{r}$ and $L$-uncoupling $L^{sr}_{ij}(R)= \langle\Psi^{sr}_i|[\hat{L_x}\pm i\hat{L_y}]/\sqrt{2}|\Psi^{sr}_j\rangle_\textbf{r}$ matrix elements between the states treated.

Since the eigenvalues of the Hamiltonian matrix depend on a relative sign of the non-diagonal matrix elements we took particularly care on the correct $R$-behaviour of the coupling functions in the entire interatomic range. At the dissociation limit the signs and normalized coefficients of the present SO matrix elements were defined to coincide with those of Ref.~\cite{Kato1993}. The correct relative sign and smooth behaviour of the CI-CPP matrix elements are guaranteed  by getting them from the same {\it ab initio} calculation in order to keep a phase of the required electronic wavefunctions unchangeable. 

All relevant CI-CPP data are picked up in the Supplementary Material~\cite{SM}. Furthermore, $A_{so}$, $\xi_{Ab0}$, $\xi_{Bb}$, $\xi_{Bc}$, and $\xi_{bc}$ SO  functions, as well as the $L$-uncoupling $L_{AB}$ and $L_{bc}$ matrix elements, which were explicitly involved in the 6$\times$6 CC molecular Hamiltonian model in Sec.~\ref{Hmodel}, are depicted in Figs.~\ref{figSOab} and~\ref{figSOemp}. It should be reminded that only $b^3\Pi_u(0_u^+)$ and $A^1\Sigma^+_u$ state PECs, along with the on-diagonal $A_{so}(R)$ and off-diagonal $\xi_{Ab0}(R)$ SO-functions (more precisely, only their corrections) have been adjusted during the global DDF procedure, while all other interatomic potentials and non-adiabatic matrix elements were constrained on their {\it ab initio} values, see Fig.~\ref{figSOemp}.

The adiabatic interatomic potentials for (1, 2)$0_u^+$, (2)$1_u$, and (1)$2_u$ states of K$_2$ corresponding to the pure Hund's "{\bf{c}}" coupling case  have been evaluated on the sparse grid in the range of $R\in [2.5,10]$~\AA~within the framework of the fully relativistic FS-RCCSD  calculations presented in Sec.~\ref{FSRCC}. The PECs for the SO-deperturbed $A0^+_u$ or $b0^+_u$ states together with the relevant off-diagonal $\xi_{Ab0}(R)$ SO matrix element were constructed during the FS-RCC calculations as well. The resulting FS-RCC point-wise functions are given in the Supplementary Material~\cite{SM}, along with the original fully relativistic PECs. 

The interatomic PECs for the excited K$_2$ states were finally constructed through the so-called "difference-based" and ground-state potentials~\cite{Zaitsevskii:05}
\begin{equation}\label{diffpecs}
   U^{*}_i = [U^{ab}_i - U^{ab}_X] + U^{emp}_X
\end{equation}
implicitly allowing to diminish the $R$-depended systematic errors in the original {\it ab initio} CI-CPP or/and FS-RCCSD PECs $U^{ab}_i(R)$. Hereafter, $U^{emp}_X(R)$ is the empirical ground-state PEC determined in the analytical form in Ref.~\cite{Tiemann2020}. 

The FS-RCC PECs of the SO-deperturbed $b^3\Pi_u(0_u^+)$ and $A^1\Sigma^+_u(0_u^+)$ states were used to get the initial sets for the non-linear parameters of the analytical DELR potentials, which were then optimized in the global DDF procedure. 

The resulting parameters of the DELR potentials defined in Sec.~\ref{analytic} and empirically adjusted during the  DDF procedure for the $A^1\Sigma^+_u$ and $b^3\Pi_u(0_u^+)$ states of K$_2$ are given in Table~\ref{DELRpar} and in Supplementary Material~\cite{SM}. The dispersion coefficient $C^{\Sigma}_3$ of the singlet $A$ state, the dissociation energy $T_{dis}\equiv T_{dis}^A$ of the $A$ state, as well as the well depths $D_e$, the equilibrium distances $R_e$ and the $\beta_i$ coefficients for both $A$ and $b$ states are considered as the adjustable fitting parameters during the least-squares processing (see Sec.~\ref{fit}), while $R_{ref}$ and $p$ parameters were kept fixed during the fit, see Table~\ref{DELRpar}. 

The fitted parameters of on-diagonal $\Delta A_{so}(R)$ and off-diagonal $\Delta \xi_{Ab0}(R)$ SO-correction functions are presented in Table~\ref{SOF} and Supplementary Material~\cite{SM}. The total resulting empirical, semi-empirical, and {\it ab initio} SO $A_{so}(R)$ and $\xi_{Ab0}(R)$ functions are compared in Fig.~\ref{figSOemp}. It is clearly seen that the present empirical SO-functions agree very well (within 0.3 - 0.5 cm$^{-1}$) with their both CI-CCP and FS-RCC {\it ab initio} counterparts, as well as with the semi-empirical SO-functions. The latter were derived in Ref.~\cite{Manaa2002} by the smooth scaling of their original {\it ab initio} MR-CI data on the appropriate experimental molecular and atomic data. Furthermore, the present empirical $A_{so}(R_e^{b0^+_u})=22.34$ and $\xi_{Ab0}(R_c^{Ab0})=13.28$ (in cm$^{-1}$) values agree very well with their previous counterparts 21.5/21.75 and 13.0/13.18 cm$^{-1}$, respectively, derived in Ref.~\cite{Ross87} and Ref.\cite{JongJMS1992} using the conventional EHA method. At the same time, both present empirical and theoretical $A_{so}$($R_e$)-values diverse from their {\it ab initio} CI-CPP-$l$ estimates~\cite{Magnier2009} by more than 5 cm$^{-1}$ (>20\%), see Table~\ref{TeRe}. It should be also noted that the reference FS-RCCSD calculation of the K atom provides the K($4p$) atomic SO constant $\xi^{K}_{so}$ = 18.85 cm$^{-1}$ which is only about 0.4 cm$^{-1}$ less than its experimental counterpart~\cite{Tiecke2019}.

The resulting empirical energies $T_e$ of $A$ and $b$ states of K$_2$ obtained by the elaborated 6$\times$6 CC DDF procedure are presented in Table~\ref{TeRe}. The present $A^1\Sigma^+_u$ state $T_e$-value is very close to $T_e$ from~\cite{Lisdat2006}, being slightly smaller than the ones given in Refs.~\cite{Manaa2002} and~\cite{Lisdat2001}, by ca 0.07 and 0.1 cm$^{-1}$, respectively. The present empirical $T_e$-value for the $b(1_u)$ state are also smaller, the respective differences being 0.3 and 0.75 cm$^{-1}$. The present empirical $R_e$-values are very close to both 4$\times$4 CC and EHA data presented in Refs.\cite{Lisdat2006, Ross87, Lisdat2001, JongJMS1992}. A comparison of the {\it ab initio} CI-CPP, CI-CPP-$l$, and FS-RCC electronic energies and equilibrium distances of the $A^1\Sigma^+_u(0_u^+)$ and $b^3\Pi_u(0_u^+,1_u,2_u)$ states of K$_2$ obtained within both Hund's "{\bf a}" and "{\bf c}" coupling cases representation of the mutually perturbed states is also given in Table~\ref{TeRe}. The present {\it ab initio} FS-RCC calculations seam to be closer to their empirical counterparts than their respective CI-CPP and CI-CPP-$l$ calculations. As can be seen from Table~\ref{TeRe}, the FS-RCC $T_e$ energies obtained for the $A^1\Sigma^+_u$ and $b^3\Pi_u(0_u^+)$ states are only by 54(<0.5\%) and 18(<0.2\%) cm$^{-1}$, respectively, lower than their empirical values. Meanwhile the relative accuracy of the CI-CPP-$l$ energies~\cite{Magnier2004, Magnier2009} is slightly higher than those of the very similar current CI-CPP estimates.
 
The empirical ground state dissociation energy $D^X_e$ given in Table~\ref{tab:Dx} was estimated from the present $T_{dis}$-value in Table~\ref{DELRpar} and the well known non-relativistic  $4p^2$P-$4s^2$S transition $\nu^K_{4p-4s}$ energy of K atom~\cite{Tiecke2019} by using the rearranged relation (\ref{Diss}): $D_e^X = T_{dis} - \nu^K_{4p-4s}$. The present $D^X_e$-value agrees within the error intervals with the empirical values reported in Refs.~\cite{Tiemann2020, Falke2006, Asen2008}. The $D_e^X$-value in Ref.\cite{Falke2006} was obtained with a help of the long-range deperturbation analysis of the highly excited rovibronic levels of the $A\sim b$ complex~\cite{Falke2007PhD}. This SO-deperturbation of the $A\sim b$ complex has been accomplished under the conventional adiabatic approximation based on the analytical PEC construction for the artificially isolated $A^1\Sigma^+_u$ state. The alternative $D_e^X$-values were obtained in Refs.\cite{Asen2008, Tiemann2020, Falke2008} from high resolution molecular spectroscopy of the ground $X$-state and scattering experiments on the ultracold potassium atoms.

Table~\ref{tab:C3tau} contains the long-range coefficients $C^{\Sigma}_3$ corresponding to the K($4s$)+K($4p$) asymptote and the radiative lifetimes $\tau^K_{4p}$ for the $^2{\rm P}_{\frac{1}{2}}$ and $^2{\rm P}_{\frac{3}{2}}$ fine components of the K($4p$) atom. The elaborated  $C^{\Sigma}_3$-value in Table ~\ref{DELRpar} was used to evaluate the present $\tau^K_{4p}$-values in accordance with Eq.(\ref{tau_K}). All data in Table~\ref{tab:C3tau} are compared with the empirical values from other sources; the present $C^{\Sigma}_3$ and $\tau$ values fit the best with the data given in Ref.~\cite{WangPRA97}. The {\it ab initio} $C_3^{\Sigma}$ and $\tau^K_{4p}$ values in Table~\ref{tab:C3tau} were obtained using the theoretical transition dipole moment $|d^K_{4p-4s}|$ = 2.918 $a.u.$ and the relativistic $\nu^{K}_{4p\;^2{\rm P}_{\frac{1}{2}}-4s\;^2{\rm S}_{\frac{1}{2}}}$ = 12979.7 and $\nu^{K}_{4p\;^2{\rm P}_{\frac{3}{2}}-4s\;^2{\rm S}_{\frac{1}{2}}}$ = 13036.3 (in cm$^{-1}$) transition energies derived within the framework of the reference FS-RCCSD-FF atomic calculations~\cite{Krumins2020}. It should be stressed that the {\it ab initio} FS-RCCSD-FF method~\cite{ZaitsevskiiTDM} provided $C_3^{\Sigma}$ and $\tau^K_{4p}$ values, which coincide within their most accurate empirical counterparts with a systematic error only about 0.5\%.

\subsection{Verification of the applied DDF procedure}\label{ver}

\subsubsection{Isotopic substitution effect}\label{ISE}

The molecular Hamiltonian construction presented in Sec.~\ref{Hmodel} depends on the reduced molecular mass $\mu$ in an explicit form, and, hence, one can expect that the empirically adjusted electronic structure parameters (excluding, perhaps, adiabatic DELR potentials of both $A$ and $b$ states) of the present 6$\times$6 CC model should be mass-invariant. To prove this assumption, we used the empirical parameters derived during the least-squares processing (see Sec.\ref{fit}) of the experimental term values belonging exclusively to the most abundant isotopologue $^{39}$K$_2$ to evaluate the rovibronic term-values and the respective multi-component (non-adiabatic) vibrational wavefunctions for other isotopologues just by substituting the appropriate molecular mass in the CC equation (\ref{CC}). 

\begin{figure}
\includegraphics[scale=0.4]{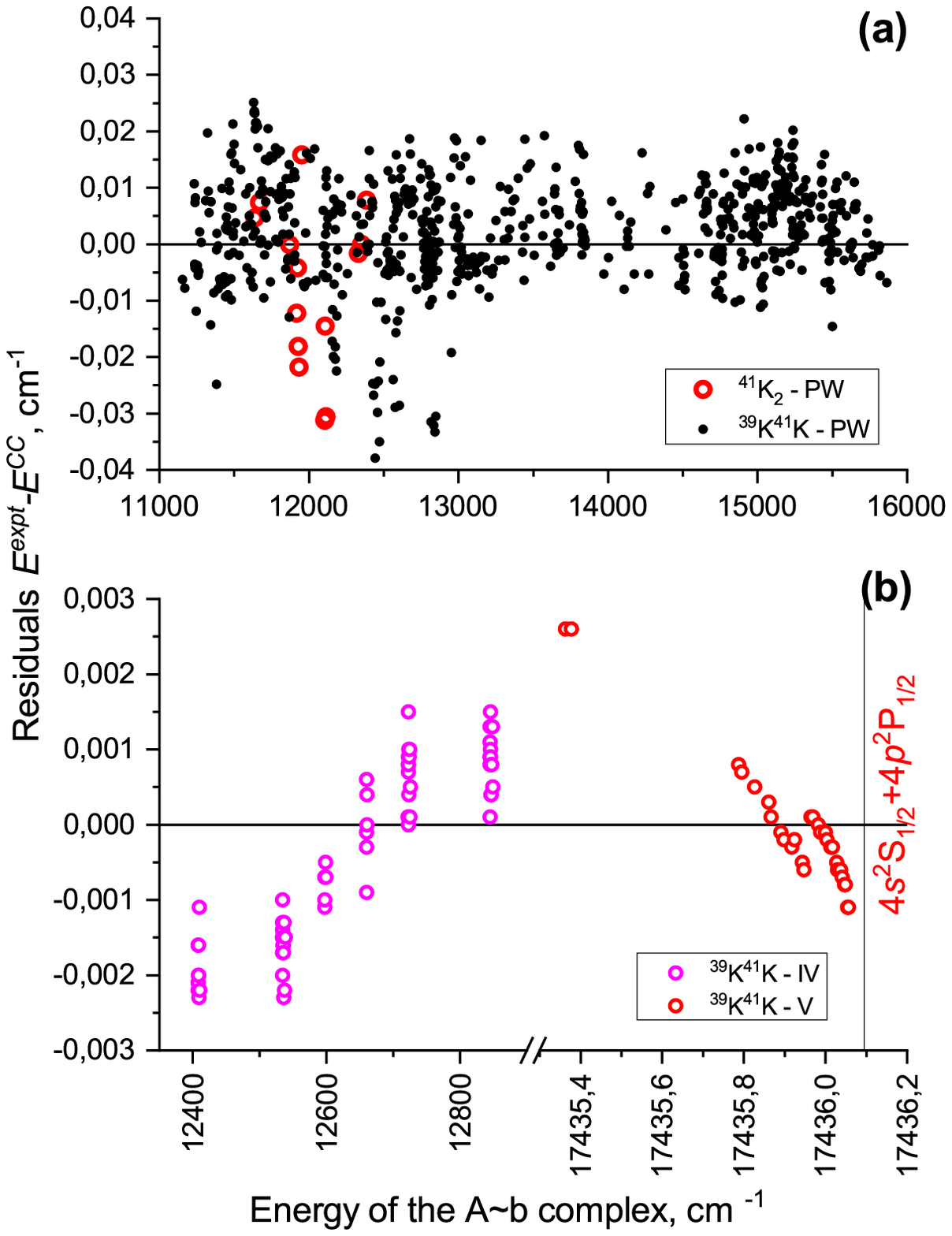}
\caption{Residuals $E^{expt}_j-E^{CC}_j$ between the experimental and calculated term values not included in the DDF procedure as dependent on the $A\sim b$ complex energy: (a) present work (PW) - black points for $^{39}$K$^{41}$K, red circles for $^{41}$K$_2$; (b) from the sub-Doppler 2s-LES-MB experiments: magenta ~\cite{Lisdat2001PhD} (set IV in Table~\ref{tab:dataset}) and red circles~\cite{Falke2007PhD} (set V in Table~\ref{tab:dataset}).} 
\label{Termvalue3941}
\end{figure}

A comparison of the term values predicted for the minor $^{39}$K$^{41}$K and $^{41}$K$_2$ isotopologues with their experimental counterparts is given in Fig.~\ref{Termvalue3941}. As can be seen the residuals between the experimental term values obtained in present work and their predicted values agree within ca. $\pm$ 0.01 cm$^{-1}$, see Fig.\ref{Termvalue3941}a, which corresponds to the experimental error $\sigma^{expt}_{i}$ = 0.015 cm$^{-1}$ in Table~\ref{tab:dataset}. However, a week systematic shift 0.003 cm$^{-1}$ (see column $Mean$) is obvious. The residuals with respect to the experimental values from ~\cite{Falke2007PhD, Lisdat2001PhD, Pazyuk2015}, see Fig.~\ref{Termvalue3941}b, are even smaller lying within ca. $\pm$ 0.002 cm$^{-1}$.

\subsubsection{Relative intensity distributions}\label{rel_int}

\begin{figure}
\includegraphics[scale=0.4]{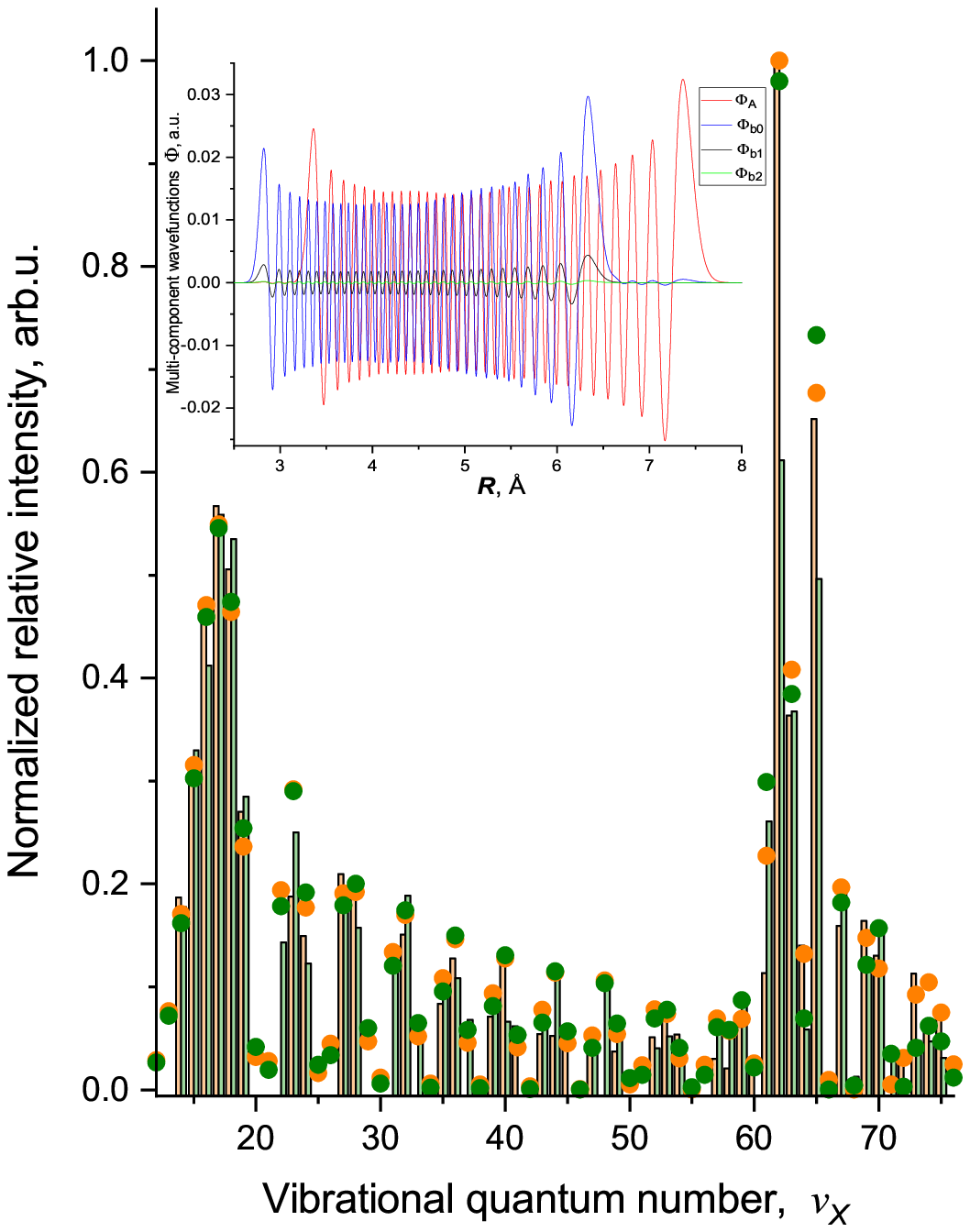}
\caption{Experimental (bars) and calculated (points) relative intensity distributions normalized to the most intensive transition in the $A\sim b \rightarrow X$ LIF $P,R$-progressions from the $E_{A\sim b}$($J^{\prime}$ = 52) = 15063.108 cm$^{-1}$ level of $^{39}$K$_2$ (green - $P$-branch, orange - $R$-branch). The fraction partition $P_i$ of $A$/$b0$/$b1$/$b2$ components is evaluated as 58/41/1/0 (in \%). The inset presents the obtained multi-component wavefunctions $\phi_i(R)$ (see text).} \label{intensities3939}
\end{figure}

\begin{figure}
\includegraphics[scale=0.6]{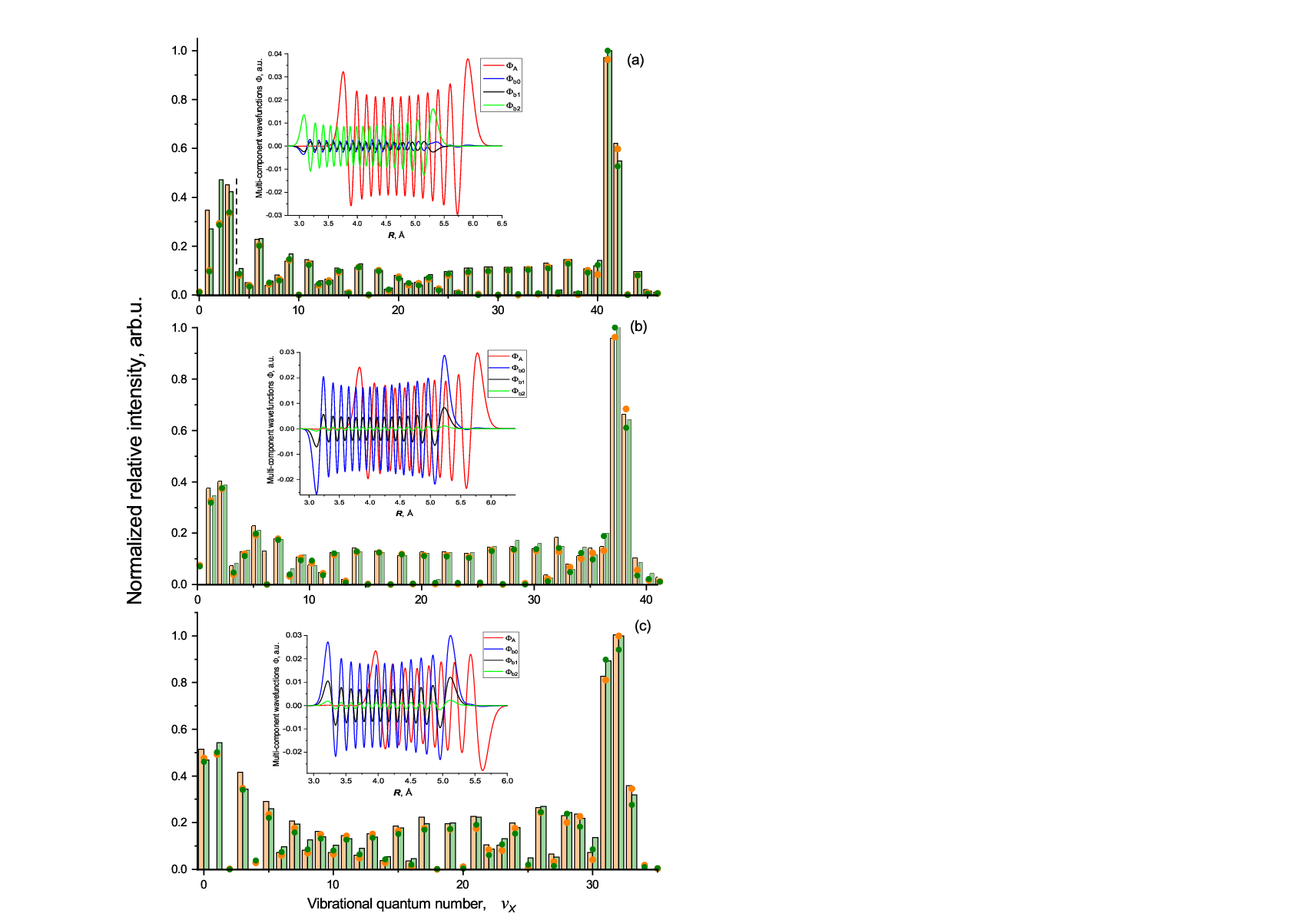}
\caption{Experimental (bars) and calculated (points) relative intensity distributions normalized to most intensive transition in the $A\sim b\rightarrow X$ LIF $P,R$-progressions (green - $P$-branch, orange - $R$-branch): (a) from the $E_{A\sim b}$($J^{\prime}$ =59) = 12853.512 cm$^{-1}$ level of $^{39}$K$^{41}$K, the fraction partition $P_i$ of $A$/$b0$/$b1$/$b2$ components is evaluated as 85/0.7/0.4/13 (in \%); (b) from the $E_{A\sim b}$($J^{\prime}$=76) = 12694.294 cm$^{-1}$ level of $^{39}$K$^{41}$K, $P_i$ is 49/47/4/0; (c) from the $E_{A\sim b}$($J^{\prime}$ = 110) = 12618.06 cm$^{-1}$ level of $^{39}$K$^{41}$K, $P_i$ is 41/50/8/0.3. The inset presents the obtained multi-component wavefunctions $\phi_i(R)$ (see text).}\label{intensities3940-1}
\end{figure}

The experimental relative intensity distribution for $A\sim b\to X$ LIF progressions originating from a particular rovibronic $J^{\prime}$-level of the $A\sim b$ complex are presented in Figs.~\ref{intensities3939} and \ref{intensities3940-1}, see bars. The intensities are normalized to the most intensive line of the LIF progression and compared with the respective theoretical calculations in accordance with Eq.(\ref{Itensinglet}). We have selected the long enough LIF progressions, which belong to different isotopologues and demonstrate the cases with different partial contributions of the $0^+_u$ component of the triplet $b$ state. As is seen, the figures demonstrate quite satisfactory agreement between measured and calculated relative intensity values for different isotopologues, see Fig.~\ref{intensities3939} for $^{39}$K$_2$ and Fig.~\ref{intensities3940-1} for $^{39}$K$^{41}$K. It is important to stress that this relates also to the cases with markedly strong fraction partition of the $b0^+_u$, $b1_u$ and $b2_u$ states, being above 40\% for $b0^+_u$ in Fig.~\ref{intensities3939} and Fig.~\ref{intensities3940-1}b,c. For the $b1_u$ it is 8\% in Fig.~\ref{intensities3940-1}c and 13\% for $b2_u$ in Fig.~\ref{intensities3940-1}a. It should be noted that the detector sensitivity drops rapidly for transition frequencies higher than 12~000 cm$^{-1}$, which may result in larger experimental uncertainties for low $v_X$, see, for instance, the transitions to $v_X$<4 in Fig.~\ref{intensities3940-1}a.

\subsubsection{Comparison with $^{39}$K$_2$ term values from other sources not included in the DDF procedure}\label{not_incl}

\begin{figure}
\includegraphics[scale=0.5]{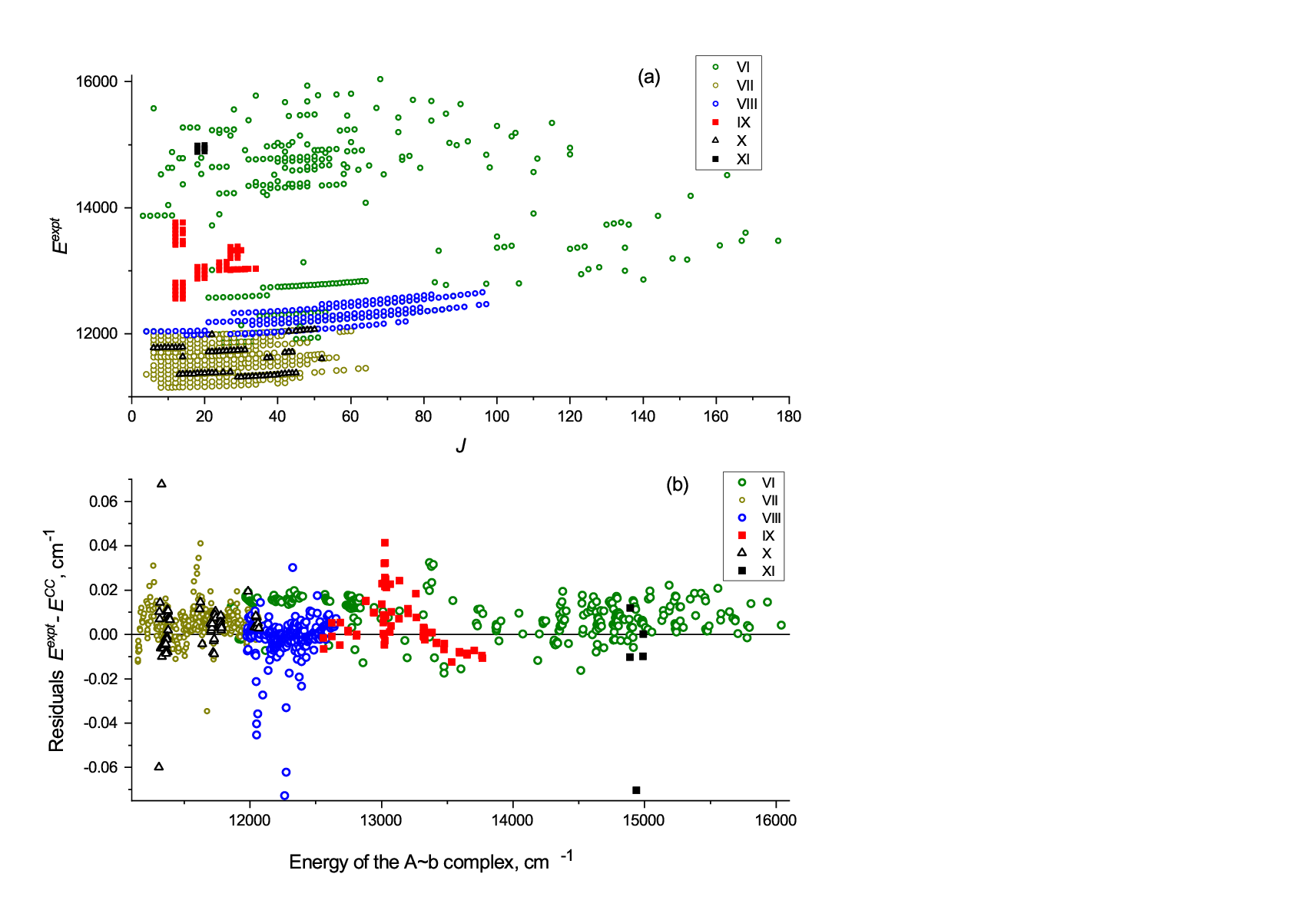}
\caption{The $^{39}$K$_{2}$ $A\sim b$ complex data from the sources not explicitly included in the DDF procedure: (a) - experimental term values; (b)  $E^{expt}_j-E^{CC}_j$ - residuals between experimental values and the values calculated in present work.}\label{Kother}
\end{figure}

Let us analyze how the resulting 6$\times$6 CC model describes the data sets in Table~\ref{tab:dataset} from VI to XI not included in the present DDF procedure, which are exposed in Fig.~\ref{Kother}a. The dispersion for majority of respective residuals, see Fig.~\ref{Kother}b, demonstrate the agreement within ca. $\pm$ 0.015 cm$^{-1}$ or even better, which agrees with the experimental uncertainties $\sigma^{expt}_{i}$ in Table~\ref{tab:dataset}. Note that the Dunham correction $Y^{X}_{00}$ has been applied (except the set VIII) to account for differing values of the ground $X$-state rovibronic energies~\cite{Heinze1987} applied to obtain VII, IX, X, and XI data sets. A small systematic shift can be noticed, see the respective values in $Mean$ column. 

Another way of verification is presented in Table~\ref{tab:termtest}, where the $E^{expt}_j$ values were obtained by us using the information given in the respective references. For Ref.~\cite{Chu2005} the  $E^{expt}_j$ values were obtained from two-step laser excitation experiment 2$^3\Pi_g\leftarrow A\sim b\leftarrow X$  (Table~\ref{tab:dataset} in Ref.~\cite{Chu2005}) by subtraction the probe laser frequencies  $\nu_{probe}$ from the energy of the 2$^3\Pi_g$ level: $E$(2$^3\Pi_g$) – $\nu_{probe}$. The term value from the one-colour PF-OODR experiment~\cite{Manaa2002} was obtained by adding laser frequency $\nu_{probe}$ = 12024.581 cm$^{-1}$ to the ground state energy level:  $E_X$($v_X$ = 1, $J^{\prime\prime}$ = 67) + $\nu_{probe}$. In the same way other presented $E^{expt}_j$-values have been obtained from the experimental data in Table~\ref{tab:dataset} of Ref.~\cite{Sherstov2007}, Table~\ref{tab:dataset} of Ref.~\cite{Asen2008}, and Table~\ref{DELRpar}, of Ref.~\cite{Falke2008}. All energies $E^{expt}_j$ presented in Table~\ref{tab:termtest} include the correction $Y^{X}_{00}$~ =~${-}$0.022 cm$^{-1}$.


\section{Summary and Conclusions}

A considerable amount of high resolution rotationally resolved Fourier Transform LIF spectra of $A \sim b \rightarrow X$ transitions of K$_2$ molecules in natural mixture of potassium vapor has been recorded and analyzed. The obtained $^{39}$K$_2$ term values were compiled with the previously published ones and applied for the direct coupled-channel (CC) deperturbation analysis of the discrete levels of the entire $A\sim b$ complex. This made it possible to cover the $A \sim b$ energy range from the very bottom of the $b^3\Pi_u$ state up to the dissociation energy of K($4s\;^2{\rm S}_{\frac{1}{2}}$)+K($4p\;^2{\rm P}_{\frac{1}{2}}$) of separated relativistic atoms. This task proposed to apply the global 6$\times$6 CC deperturbation model, which took into account explicitly both SO and electronic-rotational (Coriolis) interactions between all four $A^1\Sigma^+_u$, $b^3\Pi_u$, $c^3\Sigma^+_u$, and $B^1\Pi_u$ scalar-relativistic states converging to the common non-relativistic K($4s$)+K($4p$) dissociation limit. The explicit consideration of the higher lying $B^1\Pi_u$ and $c^3\Sigma^+_u$  states in the applied DDF procedure looks somewhat redundant, however, it guarantees the precise  simultaneous treatment all quasi-degenerate rotational $J^{\prime}>0$ levels located in vicinity of the dissociation threshold, where electronic-rotational interaction indeed is very weak but not negligible. Furthermore, the current 6$\times$6 CC deperturbation model is ready in principle to fit $0^-_u$, $1_u$, and $2_u$ terms of the $A \sim b\sim c\sim B$ complex, as well as to include the quasi-bound rovibronic levels located above the first relativistic threshold.

The initial parameters of the 6$\times$6 CC deperturbation model as a functions of the internuclear distance have been estimated in the framework of alternative scalar-relativistic and fully-relativistic {\it ab initio} electronic calculations accomplished by multi-referenced configuration-interaction and coupled-clusters methods. Among the considered {\it ab initio} methods the most robust interatomic potentials, SO coupling functions and transition dipole moments seem to be obtained in the framework of the FS-RCCSD-FF method. 

The analytical DELR interatomic potentials of the $A^1\Sigma^+_u(0^+_u)$ and $b^3\Pi_u(0^+_u)$ states along with the on-diagonal $A_{so}(R)$ and off-diagonal $\xi_{Ab0}(R)$ spin-orbit functions have been globally refined to fit the experimental $^{39}$K$_2$ data massive non-uniformly distributed between the bottom of the $b^3\Pi_u$ state and the K($4s\;^2{\rm S}_{\frac{1}{2}}$)+K($4p\;^2{\rm P}_{\frac{1}{2}}$) dissociation limit within the uncertainty of their spectroscopic measurements. The resulting parameters of the deperturbation model allowed us to predict the energies of both \textit{locally} and \textit{regularly} perturbed levels of the $A^1\Sigma^+_u$ and $b^3\Pi_u$ states with an accuracy of about 0.01 cm$^{-1}$, or better, which agrees with a usually reached experimental accuracy. Moreover, the present CC model reproduces the sub-Doppler term values derived from the 2s-LES-MB measurements with an accuracy of about 0.003-0.005 cm$^{-1}$, which is mainly limited by the uncertainty in the ground $X$-state potential.

The applied deperturbation analysis provided the refined values of empirical dissociation energy $T_{dis}$ and the long-range coefficient $C_3^{\Sigma}$ = 2$C_3^{\Pi}$, which are both related to the non-relativistic atomic threshold K($4s$)+K($4p$). The derived $T_{dis}$=17474.569(5) cm$^{-1}$ value yielded an accurate well depth $D_e$ = 4450.910(5) cm$^{-1}$ for the ground $X^1\Sigma^+_g$ state of $^{39}$K$_2$, whereas the $C_3^{\Sigma}$ = 5.501(4)$\times10^5$cm$^{-1}$\AA$^3$ value yielded the improved atomic K($4p^2$P$_{1/2;3/2}$) radiative lifetimes $\tau_{\frac{1}{2}}$=26.67(3) and $\tau_{\frac{1}{2}}$=26.32(3) $ns$. 

To clime the mass-invariant property of electronic structure parameters of the applied deperturbation model the rovibronic term values predicted for $^{39}$K$^{41}$K and $^{41}$K$_2$ isotopologues were directly compared with their experimental counterparts. A high reliability of nodal structure and fraction partition of the non-adiabatic wavefunctions of the $A\sim b$ complex was independently confirmed by a good agreement of the estimated Einstein emission coefficients with the relative intensity distributions measured for the long $A\sim b \rightarrow X^1\Sigma^+_g(v_X)$ band progressions of both $^{39}$K$_2$ and $^{39}$K$^{41}$K isotopologues.

The present analysis of the experimental data from different sources revealed a necessity in some cases to apply a systematic uniform correction most probably due to some differences of the exploited Dunham molecular constants and empirical potentials for the ground $X$-state. A further improvement of description of the entire state manifold converging to both first $4s\;^2{\rm S}_{\frac{1}{2}}+4p\;^2{\rm P}_{\frac{1}{2}}$ and second $4s\;^2{\rm S}_{\frac{1}{2}}+4p\;^2{\rm P}_{\frac{3}{2}}$ relativistic dissociation limits  would need additional empirically-based information on all $0^{\pm}_u$, $1^{\pm}_u$ and $2^{\pm}_u$ components belonging to $A$, $b$, $B$ and $c$ states of K$_2$. 

\begin{acknowledgments}

Riga team acknowledges the support from the University of Latvia Faculty of Physics, Mathematics an Optometry, project: "A detailed description of the strongly mixed first excited states of the K$_2$ molecule: challenges and solutions", the Laserlab-Europe EU-H2020 project No 871124 and from the University of Latvia Base Funding No A5-AZ27. Moscow team is grateful for the support by the Russian Government Budget (section 0110; Project Nos. 121031300173-2 and 121031300176-3). 
\end{acknowledgments}

\section*{Data Availability Statement}

The data that support the findings of this study are available within the article [and its supplementary material~\cite{SM}].

 AIP Publishing believes that all data sets underlying the conclusions of the paper should be available to readers. Authors are encouraged to deposit their datasets in publicly available repositories or present them in the main manuscript. All research articles must include a data availability statement stating where the data can be found. In this section, authors should add the respective statement from the chart below based on the availability of data in their paper.

\begin{center}
\renewcommand\arraystretch{1.2}
\begin{tabular}{| >{\raggedright\arraybackslash}p{0.3\linewidth} | >{\raggedright\arraybackslash}p{0.65\linewidth} |}
\hline
\textbf{AVAILABILITY OF DATA} & \textbf{STATEMENT OF DATA AVAILABILITY}\\  
\hline
Data available on request from the authors
&
The data that support the findings of this study are available from the corresponding author upon reasonable request.
\\\hline
Data available in article or supplementary material~\cite{SM}
&
The data that support the findings of this study are available within the article [and its supplementary material~\cite{SM}].
\\\hline
Data openly available in a public repository that issues datasets with DOIs
&
The data that support the findings of this study are openly available in [repository name] at http://doi.org/[doi], reference number [reference number].
\\\hline
Data openly available in a public repository that does not issue DOIs
&
The data that support the findings of this study are openly available in [repository name], reference number [reference number].
\\\hline
Data sharing not applicable – no new data generated
&
Data sharing is not applicable to this article as no new data were created or analyzed in this study.
\\\hline
Data generated at a central, large scale facility
&
Raw data were generated at the [facility name] large scale facility. Derived data supporting the findings of this study are available from the corresponding author upon reasonable request.
\\\hline
Embargo on data due to commercial restrictions
&
The data that support the findings will be available in [repository name] at [DOI link] following an embargo from the date of publication to allow for commercialization of research findings.
\\\hline
Data available on request due to privacy/ethical restrictions
&
The data that support the findings of this study are available on request from the corresponding author. The data are not publicly available due [state restrictions such as privacy or ethical restrictions].
\\\hline
Data subject to third party restrictions
&
The data that support the findings of this study are available from [third party]. Restrictions apply to the availability of these data, which were used under license for this study. Data are available from the authors upon reasonable request and with the permission of [third party].
\\\hline
\end{tabular}
\end{center}


\end{document}